\documentclass[letterpaper,12pt]{article}
\usepackage[utf8]{inputenc}
\usepackage{amsmath}

\usepackage{bm}

\usepackage[plain,noend]{algorithm2e}

\usepackage[letterpaper,margin=1in]{geometry}
\usepackage{natbib}
\usepackage{setspace}
\usepackage{multicol}
\usepackage{fancyhdr}
\usepackage{float}
\usepackage{pgf}
\usepackage{tikz}
\usepackage{amsthm}
\usepackage{amsmath}
\usepackage{setspace}
\usepackage{lscape}
\usepackage{enumerate}
\usepackage{fancyvrb}
\usepackage{parskip}
\usepackage{enumerate}
\usepackage{amsfonts}
\usepackage{framed}
\usepackage{mathabx}
\usepackage{graphicx}
\usepackage{times}
\usepackage{makecell}
\usepackage{booktabs}
\usepackage{caption}
\usepackage{subcaption}
\usepackage{multirow}

\newtheorem{theorem}{Theorem}[section]

\newtheorem{corollary}{Corollary}[theorem]

\theoremstyle{definition}

\usetikzlibrary{shapes,shapes.geometric,decorations,arrows,calc,arrows.meta,fit,shapes.arrows,shapes.multipart,decorations.pathmorphing, positioning}
\tikzset{
    -Latex,auto,node distance =1 cm and 1 cm,semithick,
    state/.style ={ellipse, draw, minimum width = 0.7 cm},
    point/.style = {circle, draw, inner sep=0.04cm,fill,node contents={}},
    bidirected/.style={Latex-Latex,dashed},
    el/.style = {inner sep=2pt, align=left, sloped}
}
\DeclareMathOperator\supp{supp}

\DeclareMathOperator{\EX}{\mathbb{E}}%
\date{\today}

\begin{document}

\title{Extrapolating Single-Treatment Effects Out of Factorial Experiments}

\author{Guilherme Duarte}

\maketitle

\begin{abstract}
Despite their cost, randomized controlled trials (RCTs) 
are widely regarded as gold-standard evidence in disciplines ranging from 
social science to medicine. In recent decades, researchers 
have increasingly sought to reduce the resource burden of 
repeated RCTs with \emph{factorial designs} that simultaneously 
test multiple hypotheses, e.g. experiments that evaluate the 
effects of many medications or products simultaneously. This paper shows 
that when multiple interventions are randomized in experiments, the
extrapolation of single-treatment effects outside the experimental setting 
 is generally not identified without strong assumptions, even under otherwise realistic conditions.
This happens because single-treatment effects involve a counterfactual 
world with a single focal intervention, allowing other variables to 
take their natural values (which may be confounded or modified by the 
focal intervention). In contrast, observational studies and factorial 
experiments provide information about potential-outcome distributions 
with zero and multiple interventions, respectively.  Therefore, unobserved confounding 
affects not only the extrapolation from observational to experimental settings
but also the extrapolation from factorial experiments to single-treatment settings.
In this paper, I formalize sufficient conditions for the identifiability of 
those isolated quantities. Researchers who rely on this type of design 
need to justify either functional and/or structural assumptions. 
  Finally, I develop nonparametric sharp bounds---i.e., 
maximally informative best-/worst-case estimates consistent with 
limited RCT data---that show when extrapolations about effect signs 
are empirically justified. These new results are illustrated with simulated data.
\end{abstract}


\newpage 

\doublespacing

\section{Introduction}

Randomized controlled trials are now widely considered the gold standard in empirical research. Researchers have used them to assess effects ranging from medication to economic development and discrimination. But their logistical complexity and cost are compounded by scientists' common
  need to test many hypotheses.  As a result, scholars turn to factorial experiments, which evaluate multiple
  hypotheses simultaneously.
  Papers using this general design number in the tens of 
  thousands\footnote{Recent theoretical results were reported by \citet{dasgupta2015causal, lee2020general, lee2020identification, egami2018causal, de2022improving,  zhao2022regression,blackwell2023noncompliance, pashley2023causal}. } and variants like conjoint analysis are an increasingly dominant
  approach in market research and survey experiments.  Among the advantages of factorial designs, 
  statisticians believe they provide \emph{greater efficiency} because they allow the evaluation of factors with 
  only one-quarter of the data that otherwise would be necessary, and \emph{greater comprehensiveness}, as not only 
  a few factors are analyzed but also their interactions \citep[p.~113]{fisher1949design}.
  Recently, Nobel Laureates in Economics
  Esther Duflo and Michael Kremer argue, ``because they significantly reduce costs, cross-cutting different treatments [i.e., factorial design] has proved 
  very important in allowing for the recent wave of randomized evaluations in development economics" 
  \citep{duflo2007using}. Randomization is often understood to allow scientists to infer ``model-free," or nonparametric, estimates of average treatment effects.

However, factorial experiments do not generally support extrapolation to single-treatment effects without additional assumptions.
  For instance, while the average factorial effect for the target population is identifiable, a corresponding average single-treatment effect is 
  only identifiable under additional assumptions.   
  Intuitively, this is because factorial experiments 
  present a setting where multiple treatments are implemented at the same time. 
  However, when researchers are interested in single-treatment quantities, they are devising 
  environments where one treatment is manipulated but 
  all the others might be confounded with the outcome, 
  i.e. other variables might be causing the treatment variable and the outcome at the same time. 
  As those cases can only be assessed in trials with single treatments, 
  any extrapolation requires additional identifying assumptions; in the fully nonparametric case, 
  identification generally fails.

  Consider the following example. Suppose one is investigating individuals admitted to a hospital with symptoms of COVID-19. 
  Let a binary $Y$ denote survival after 2 months. 
  Assume $A$ denotes drug A, 
  a drug often employed to treat lupus, and $B$ denotes admission or not to the Intensive Care Unit.  
  Both are binary variables. 
  One would like to know not only the joint effect of the two treatments, 
  $A$ and $B$, on the outcome $Y$ but also their effects in isolation, 
  for instance, the single-treatment effect of $A$ on $Y$ or $B$ on $Y$. 
  If one designs a study, there are 9 cases we could consider: 
  1) 4 groups with 2-treatment interventions, A=B=1, A=0/B=1, A=1/B=0, and A=B=0; 
  2) 4 groups with 1-treatment interventions, A=0, A=1, B=0, B=1; 
  3) 1 group with no intervention, i.e. observational data.\footnote{The observational group is different from the 
  treatment group with A=0;B=0, because in the observational group, individuals might eventually 
  receive $A$ and $B$ treatments given realizations of unobserved confounders. They are only the same in the particular situation 
  where naturally no individual receives any treatment, which is the case of some program evaluations.} 
  All those cases are represented in Figure~\ref{fig:typesdata}. The first case is exactly the 
  case of a factorial experiment, while the second case is a single-armed randomized controlled trial (RCT), and the third represents observational data. Notice that observational data is not the same thing as the case $A=0; B=0$ in the factorial experiment, because $A$ and $B$ might be confounded with $Y$ in the no-treatment group. Indeed those cases are not even similar to groups $B=0$ and $A=0$ 
  in the factorial RCT, because if one intervenes on $A$, $B$ might still be confounded, and vice-versa, if one intervenes on $B$.
  Therefore, if one intends to draw inferences on the average treatment effect of $A$ on $Y$, one might use two of the groups enumerated in Figure \ref{fig:typesdata}-b (single-armed RCT), yet if they only collect data concerning the groups in Figure \ref{fig:typesdata}-a (factorial experiment), this inference is not directly supported unless one is willing to extrapolate from the factorial experiment to the single-armed one. 
  
  \begin{figure}
      \centering
      \begin{tikzpicture}
          \draw (-7,-1) rectangle (7,5); 
          \node (2treattext) at (-4.3,4) {a) 2-Treatment groups};
          \node (2treattext) at (-4.3,3.5) {(Factorial experiment)};
            \node[rectangle,draw] (2treat1) at (-4.3,2.5) {$A=1;B=1$};
            \node[rectangle,draw] (2treat2) at (-4.3,1.75) {$A=1;B=0$};
            \node[rectangle,draw] (2treat3) at (-4.3,1) {$A=0;B=1$};
            \node[rectangle,draw] (2treat4) at (-4.3,0.25) {$A=0;B=0$};
          \node (1treattext) at (0,4) {b) 1-Treatment groups};
                    \node (1treattext) at (0,3.5) {(Single-treatment RCT)};
            \node[rectangle,draw] (1treat1) at (0,2.5) {$A=1$};
            \node[rectangle,draw] (1treat2) at (0,1.75) {$A=0$};
            \node[rectangle,draw] (1treat3) at (0,1) {$B=1$};
            \node[rectangle,draw] (1treat4) at (0,0.25) {$B=0$};
        \node (notreatment) at (4.5,4) {c) No-treatment group };
        \node (notreatment) at (4.5,3.5) {(Observational data)};
            \node[rectangle,draw] (notreat1) at (4.5,2) {No intervention};
      \end{tikzpicture}
\caption{Study of individuals affected by COVID-19 in a hospital. 
The two binary treatments $A$ and $B$ represent respectively the 
drug A and admission to the Intensive Care Unit (ICU). 
All the possibilities of treatment and control (interventions) are enumerated. 
In column (a), a 2x2 factorial experiment, 
there are four possibilities of interventions at the same time on both $A$ and $B$. 
In column (b), 
there are four cases of single-treatment interventions on $A$ or $B$, which constitute a single-armed RCT. 
Finally, in column (c), there is only one case when there is no active intervention. 
This is the case of observational data. 
}   
      \label{fig:typesdata}
  \end{figure}

  To make things more concrete, consider a setting with incomplete confounded data. 
  For example, suppose one only has observational data and there is no experimental data. 
  In such a world, without interventions, both drug A and ICU admission are confounded 
  with $Y$ by unobserved confounders $U_{AY}$ and $U_{BY}$. 
  This situation is represented by DAG (c) in Figure~\ref{fig_example}. 
  Suppose one is interested in estimating the average treatment effect of 
  drug A on COVID-19 survival. The only available data 
is observational, and possible confounders are not observed, 
  it is not identifiable to extrapolate from (c) to (b) without strong assumptions. 
In other words, the only way we could estimate the effect of $A$ on $Y$, in this case, 
would be by implementing a single-treatment RCT or trying to observe the confounders.
 Otherwise, any attempt to answer a causal question about Figure~\ref{fig_example}-b 
  using data on Figure~\ref{fig_example}-c would not be identifiable. 
 This is the classical omitted variable bias, the known fact that an 
 attempt to estimate causal effects from observational data with unobserved confounders will be biased.

  The converse problem of trying to answer an observational question using a single-treatment RCT, 
  while not widely considered, 
  is not identifiable either in this case. 
  For example, suppose one runs an RCT 
  and assigns people to two groups where they received drug A 
  ($A=1$) or not ($A=0$). 
  In this case, one can answer questions regarding the average treatment effect of $A$ on $Y$.
   But if one intends to know observational facts about the incidence of the 
   disease in the population, for instance, what percentage 
   of COVID-19 patients survived in one hospital, 
   the experimental data would not be sufficient, or the survival rate for those who were admitted to the ICU by instruction of 
   the physician. 
   In other words, it is not identifiable to extrapolate from the single-treatment 
   experiment represented in Figure~\ref{fig_example}-b, 
   where there is intervention on $A$, and $B$ remains confounded with $Y$, 
   to the observational world represented in Figure~\ref{fig_example}-c, 
   where both $A$ and $B$ are confounded with $Y$.

  Finally, suppose that rather than implementing an RCT 
  with interventions only on $A$, one carries out a factorial experiment 
  and now has four groups with interventions on $A$ and $B$ at the same time. 
  That is a common scenario for researchers who are not exactly interested in the effect 
  of those treatments at the same time but implement them anyway for 
  for cost-saving reasons.
  The problem now has changed to answering questions about single-treatment effects 
  of $A$ (Figure~\ref{fig_example}-b), using the factorial data (Figure~\ref{fig_example}-a). 
  And the answer, analogously to the previous case  -- as shown below -- is generally negative. 
  This result might be puzzling, given that we indeed had interventions on $A$ in the factorial 
  experiment, but to answer the causal question about the single-treatment effect of $A$, 
  one is implicitly assuming a world where $B$ remains confounded with $Y$. 
  This result becomes still more puzzling when one realizes that even if they possessed 
  at the same time factorial-experimental and observational data, 
  the single-treatment effect of $A$ would not be point-identifiable without other assumptions, 
  as shown below. This result suggests one should be aware of potential problems of 
  answering causal questions through cross-counterfactual world data.

  \begin{figure}
      \begin{tikzpicture}
\begin{scope}[xshift=-5cm]
     \centering
    \node[draw, rectangle]  (a) at (0,0) {$A=a$};
    \node[state]  (y) [below right =of a] {$Y$};
    \node[draw, rectangle]  (b) [below left =of y] {$B=b$};
    \node (uay) [above =of y] {$U_{AY}$};
    \node (uby) [below =of y] {$U_{BY}$};
    \path (a) edge (y);
    \path (b) edge (y);
    \path[dashed] (uay) edge (y);
    \path[dashed] (uby) edge (y);
            \node[anchor=center] at (1, -4.5) {(a) 2-treatment ($A$ and $B$)};
\end{scope}
\begin{scope}[xshift=0cm]
     \centering
    \node[draw, rectangle]  (a) at (0,0) {$A=a$};
    \node[state]  (y) [below right =of a] {$Y$};
    \node[state]  (b) [below left =of y] {$B$};
    \node (uay) [above =of y] {$U_{AY}$};
    \node (uby) [below =of y] {$U_{BY}$};
    \path (a) edge (y);
    \path (b) edge (y);
    \path[dashed] (uay) edge (y);
    \path[dashed] (uby) edge (y);
    \path[dashed] (uby) edge (b);
            \node[anchor=center] at (1, -4.5) {(b) 1-treatment (only $A$)};
\end{scope}
\begin{scope}[xshift=5cm]
\centering
    \node[state]  (a) at (0,0) {$A$};
    \node[state]  (y) [below right =of a] {$Y$};
    \node[state] (b) [below left =of y] {$B$};
    \node (uay) [above =of y] {$U_{AY}$};
    \node (uby) [below =of y] {$U_{BY}$};
    \path (a) edge (y);
    \path (b) edge (y);
    \path[dashed] (uay) edge (y);
    \path[dashed] (uby) edge (y);
    \path[dashed] (uay) edge (a);
    \path[dashed] (uby) edge (b);
        \node[anchor=center] at (1, -4.5) {(C) No treatment};
\end{scope}
\end{tikzpicture}
 \caption{Graphs representing experimental/observational settings of 
 a study trying to assess the effect of drug A ($A$) on COVID-19 survival ($Y$). 
 $B$ denotes admission or not to the ICU. In the graphs, direct arrows denote causality direction. 
 $U_{AY}$ and $U_{BY}$ are potential unobserved variables, 
 and for that reason, we employed dashed arrows. Any direct effect of $A$ on $B$ or vice-versa is not considered here (no mediation).
 Variables in boxes denote interventions, for example, 
 fixing the value $a$ to the variable $A$ by an intervention. 
 When an intervention happens, any arrow points to those intervened variables 
 are immediately truncated. I consider three cases here. In case \textbf{(a)}, 
 representing a factorial experiment, there are two interventions 
 on $B$ and $Y$, so confounding between $A$ and $Y$ and $B$ and $Y$ are removed. 
 In case \textbf{(b)}, a single-treatment RCT, there is only one intervention setting $A$ to a. 
 Finally, case \textbf{(c)} represents an observational setting,  
 and both $A$ and $B$ are confounded with $Y$. Without major assumptions, 
 extrapolations using data from one case to answer questions about other cases are not supported. }
     \label{fig_example}
\end{figure}

In this paper, I assess when extrapolation from one case 
to another with different quantities of intervened variables is 
justified, and I develop a principled approach to estimate 
best- and worst-case scenarios for any estimand of interest in 
nonparametric settings. The approach consists of the 
following steps. First, researchers should specify the different 
worlds they are studying: observational, 1-treatment, ..., and 
n-treatments as well as the causal structure they are willing to justify. 
Secondly, they consider potential reasonable functional assumptions 
for the scenario, for instance, additive separability, 
monotonicity, or linearity. Finally, they are able to derive 
point- or partial-identification results.

The problem of generalizability from experimental settings to
target populations has a long history in statistics and
causal inference.
\citet{fisher1949design} recognized those concerns  
and emphasized 
that treatment comparisons obtained under particular experimental 
conditions need not extend to other settings, and argued that 
treatment–environment interactions are the appropriate objects for 
assessing the stability of experimental conclusions. 
Within Fisher’s framework, however, the generalizability problem was 
understood as being resolved once these interactions were estimated: 
given sufficient factorial variation across environments, 
treatment effects were taken to be fully characterized by their interaction 
structure. 
When reformulated in modern causal terms, this view is incomplete. 
Even with complete knowledge of treatment–environment 
interactions on the experimental support, 
there is no implied extrapolation from those environments to a target population, 
and hence no uniquely defined single-treatment effect for real-world deployment.\footnote{Knowing all 
interactions tells us how treatments behave where we experimented, 
but not how to aggregate or transport those effects to the 
environment in which a policy is actually applied.} 
By making the target estimand explicit, modern causal formalizations reveal that the 
absence of an extrapolation rule constitutes an identification problem rather than 
an estimation problem. The contribution of this paper is to address this gap by 
formalizing generalizability in causal terms and by providing partial 
identification bounds for target-population treatment effects under transparent and weak assumptions.

This paper has the following structure. 
In section 2, I illustrate the problem by elaborating on an experiment with two treatments. 
I show how an estimated causal effect can be biased if some of those assumptions fail.   
I demonstrate the types of assumptions required for the identification of single-treatment 
effects using factorial experiments, then show how current approaches are insufficient for 
identification when these are violated. In section 3, I derive new partial-identification 
results to address these shortcomings. 
Finally, in section 4, I apply these techniques to two simulated examples and demonstrate 
how they can be extended to sensitivity analyses that probe how results vary depending on key unmeasured quantities.

\section{Extrapolation problem}

\subsection{Non-identifiability result}

Here I study whether single-treatment effects can be identified when there is an 
arbitrary set of observational and experimental data and a specific causal query.  
In other words, one researcher might have available information on a factorial experiment, 
and maybe observational data as well, and they want to know whether
the single-treatment effect can be extrapolated from those data. 
The answer is generally negative, if no strong assumptions are made, as I show below.
On the other hand, if one is willing to make strong structural or functional assumptions, 
then identification might be possible.

For the purpose of notation, we use 
uppercase letters ($A, B, Y$) to represent random variables and lowercase 
letters ($a, b, y$) to represent their instantiations.  
Sometimes, this researcher might have observational data as well, 
denoted here by $\Pr(Y, A, B)$. In the paper, I will be assuming the 
standard regular assumptions of SUTVA and positivity \citep{imbens2015causal}. 
The notation of potential outcomes is used throughout the paper. $Y_{a, b}$ 
denotes the potential outcome $Y$ when treatments $A$ and $B$ are set to $a$ and $b$, respectively. 
Similarly, $Y_a$ denotes the potential outcome $Y$ when treatment $A$ is set to $a$, 
and $B$ takes its natural value.  
The problem here is whether it is possible to identify single-treatment 
quantities using these data. In other words, one desires to know 
whether it is possible to identify $\Pr(Y_a)$, using information on $\Pr(Y_{a, b})$ and $\Pr(Y, A, B)$.

\begin{figure}
    \caption{A nonparametric structural graph illustrating a scenario where both $A$ and $B$ cause $Y$. All three variables, $A$, $B$, and $Y$, are influenced by unobserved confounders simultaneously. With the exception of cases where structural assumptions are explicitly relaxed, all the results presented in the paper assume this graph.}
        \label{fig:nonparamgraph}
    \centering
\begin{tikzpicture}
    \begin{scope}[xshift=-2cm]
  \node[state]  (b) at (0,0) {$B$};
    \node[state]  (y) [below right =of b] {$Y$};
    \node[state]  (a) [below left =of b] {$A$};
    \path (a) edge (y);
    \path (b) edge (y);
    \path[bidirected] (a) edge[bend right=30] (y);
    \path[bidirected] (b) edge[bend left=30] (y);
    \path[bidirected] (a) edge[bend left=30] (b);
    \end{scope}
\end{tikzpicture}
\end{figure}

In the context of the nonparametric scenario depicted in Figure~\ref{fig:nonparamgraph}, 
identification does not generally hold. 
Factorial experiments, whether conducted alone or alongside observational data, 
are insufficient for addressing questions regarding single-treatment effects. 
Demonstrating non-identifiability necessitates the construction of counterexamples 
wherein at least two distinct models yield identical data but possess estimands with differing values. 
In the appendix, I provide such counterexamples, illustrating this outcome both in a 
general setting and under the assumption of monotonic effects for both treatments. 
Intuitively, this arises because inquiries into a causal domain involving only one 
intervention cannot be directly resolved using information from causal domains 
featuring multiple interventions or none at all. Formally, this conclusion is derived from the following theorem:

\begin{theorem}
  Consider treatments $A$ and $B$, along with an outcome $Y$. Assume that both $A$ and $B$ independently cause $Y$, without causing each other. Additionally, suppose the effects of both $A$ and $B$ on $Y$ are monotonic, meaning $\Pr(Y_{a_0}=y_1, Y_{a_1}=y_0) = \Pr(Y_{b_0}=y_1, Y_{b_1}=y_0) = 0$. Under these conditions, without imposing any other assumptions, such as absence of confounding between $A$, $B$, and $Y$, the probability distribution $\Pr(Y_a)$, representing the outcome under different levels of treatment $A$, is not identifiable solely from the joint distribution $\Pr(Y_{a, b})$ and $\Pr(Y, A, B)$. Furthermore, the ATE, defined as the difference in probabilities between different treatment levels, $\Pr(Y_a = y) - \Pr(Y_{a'} = y)$, remains unidentifiable. Finally, these quantities are not identifiable even if we relax the assumption of monotonicity or if we only have observational or factorial data alone.
\end{theorem}
\begin{proof}
See the proof in Theorem~\ref{theorem:nointeractmono} (in the appendix).    
\end{proof}

This finding is sufficient to establish the non-identifiability of the estimand given the problem. Consequently, attempts to estimate the single-treatment quantity using data from two treatments and observational data without substantial assumptions can yield biased results. I term this bias F-Bias.

\begin{corollary}[F-bias]
Consider an estimand of interest $\Pr(Y_{a} = y)$. Let $\hat\theta$ denote an estimator, constructed as a function of other estimators for $\Pr(Y_{a,b} = y)$ and $\Pr(y,a,b)$. Almost surely $\hat\theta$ will exhibit bias, given the non-identification result. This bias is referred to as F-bias.
\end{corollary}

The presence of $F$-bias raises the question of whether other functional 
assumptions, besides monotonicity, would lead to point-identification 
in those settings. The answer is affirmative, as the next 
subsection will demonstrate regarding additive separability (no interactions). 
However, these scenarios require both factorial and observational data. 
While the ATE is point-identifiable, 
the expected value of isolated single-treatment estimands, 
such as $\EX(Y_a)$, is not. Formal results are included 
in Theorem~\ref{theorem:nointeractmono}. Furthermore, the identification of 
the ATE is no longer valid if either observational or factorial 
data are absent (Theorem~\ref{theorem:nointeractonlyfact}). Therefore, 
these scenarios will typically cause $F$-bias.

It should be emphasized that the problem here is one of transportability (of extrapolation). 
For the population defined in the experimental design---which we denote by $Y^F_{a}$ to 
distinguish it from the population of interest--- the quantity $\Pr(Y^F_{a})$ is
identifiable from the experimental data $\Pr(Y^F_{a, b})$ by the law of total probability:
$\Pr(Y^F_{a}) = \sum_b \Pr(Y^F_{a, b}) \Pr(B^F=b)$.
However, the problem here is to identify $\Pr(Y_{a})$ for the target population, 
which might be different from the experimental one.
In other words, the problem is to extrapolate from $\Pr(Y^F_{a, b})$ and $\Pr(Y, A, B)$ to $\Pr(Y_{a})$.
A common practice in this scenario is to state that $\Pr(Y^F_{a})$ is just a second-best estimand, 
to recognize that it lacks external validity, and to hope that it is close enough to $\Pr(Y_{a})$.
However, as I show, even under otherwise realistic conditions,
this extrapolation is not generally supported without strong assumptions. In this paper, 
we provide precise conditions under which this extrapolation is justified.

\subsection{Sufficient parametric assumptions for point identifiability}

If one is willing to make parametric assumptions, 
then single-treatment quantities can be recovered 
using factorial and observational data. 
For instance,  when the data-generating process 
satisfies linearity with 
interactions, $y_i = \beta_0 + \beta_1 a_i + \beta_2 b_i + \beta_3 a_i b_i + u_i$, it 
can be shown that (proof in Theorem~\ref{theorem:lininteract}): 
\begin{equation}
    \label{equation:popbfactorial}
    \EX(Y_a) = \sum_{b \ \in \ \supp(B)} \EX(Y_{a,b})\Pr(B = b)
\end{equation}
That implies that the ATE is identifiable by:
\begin{equation}
    \label{equation:pcme}
    ATE = \sum_{b \ \in \ \supp(B)} (\EX(Y_{a,b}) - \EX(Y_{a',b}))\Pr(B = b),
\end{equation} where $P(B=b)$ is the population distribution of $B$. This linearity is not mere 
linearity of Conditional Expectation Function (CEF), but rather linearity of the real-world model,
which is a stronger assumption. 

This estimand remains extremely relevant in the 
conjoint experiment literature, as it aligns with the Average Marginal Component 
Effect (AMCE) described by \citet{hainmueller2014causal}, 
where factorial quantities are weighted by the population 
distribution of $B$. That case was also studied by \citet{zhao2022regression}. In other words, 
the ATE is recoverable by weighting the factorial effects with the population distribution of $B$, 
which is exactly the AMCE.
Notice, however, that $P(B=b)$ is the population distribution of $B$, not merely the 
uniform distribution of $B$ in the factorial experiment. Although the equation appears 
just a mere application of the law of total probability, in fact, it requires 
data fusion methods to estimate it, as $P(B=b)$ is not directly observable in the factorial experiment. 
This solution was initially proposed in \citet{de2022improving} under the name Population 
Average Marginal Component (pAMCE).

On the other hand, if the linearity form assumption is too restrictive, 
it is still possible to identify their ATE
by making the additive separability (AS) assumption. 
This assumption posits that $y_i = f(a_i, u_i) + g(b_i, u_i)$, 
where $f$ and $g$ are arbitrary functions. 
Since this functional form implies $\EX(Y_{a, b}) - \EX(Y_{a', b}) = f(a, u_i) - f(a', u_i)$ 
is constant for any $b$, the assumption is also referred to as ``no interaction'' or ``homogeneity.'' 
This implies that the ATE is identifiable through:

\begin{equation*}
\EX(Y_{a} - Y_{a'}) = \EX(f(a_1,u_i)) - \EX(f(a_0, u_i)) = \EX(Y_{a,b} - Y_{a',b}), \quad \forall b
\end{equation*}

and

\begin{equation}
\label{equation:uamce}
\sum_b (\EX(Y_{a_1, b}) - \EX(Y_{a_0, b})) \cdot h(b), \quad \text{where $h(b)$ is the uniform p.m.f. over $B$}.
\end{equation}

In other words, under the AS assumption, the single-treatment ATE can be recovered 
using either any fixed level of $B$ or by averaging over $B$ with a uniform distribution. That suggests that 
AS or no interaction is sufficient for external validity of the ATE from factorial experiments to single-treatment settings.
The formal proofs supporting this assertion are provided in Theorem~\ref{theorem:as} in the appendix.

It is important to emphasize that although the ATE may be recovered,  
the AS assumption does not 
guarantee the identifiability of single-treatment point exposures, 
even in the presence of both factorial and observational data. 
In other words, despite $\EX(Y_{a} - Y_{a'})$ being identifiable, 
$\EX(Y_{a})$ itself remains unidentifiable.
This outcome is demonstrated in Theorem~\ref{theorem:nointeractmono}. 
This situation arises because when computing the ATE, 
the constant component implied by the absence of interaction is subtracted from itself, 
but that does not happen when computing the simple expected outcomes.

Lastly, if one is willing to make both the AS assumption and linearity 
(non-interactive linearity), 
expressed as $y_i = \beta_0 + \beta_1 a_i + \beta b_i + u_i$, 
then one obtain the benefits of both assumptions. 
As a result, 
single-treatment point exposures as well as the 
ATE become point identifiable. This conclusion is concisely encapsulated in Corollary~\ref{corollary:linearity}.

In summary, adopting the assumptions of AS and linear interactions may facilitate the identification of 
single-treatment point exposures or risk differences. 
However, these assumptions should not be considered innocuous. 
For instance, 
assuming AS  implies the absence of interaction between treatments, 
suggesting that the effect of one treatment 
remains unchanged in the presence of another—a potentially 
overly restrictive assumption in certain scenarios. 
In Section 4, I will present a method for conducting sensitivity 
analysis when restricting the number of interactive units in the experiment. 
Furthermore, the assumption of linearity may be 
overly constraining.
Particularly when dealing with binary or discrete outcomes, 
assuming linearity could result in overly restrictive models, 
effectively constraining the analysis within vector spaces over fields $F_{0,1}$. 
Nevertheless, these parametric assumptions are merely illustrative, 
and researchers are not precluded from justifying other, less stringent 
structural assumptions that are sufficient for identification. All these 
results are summarized in Table~\ref{table:idresults}.

\begin{table}[h]
\centering
\caption{Point-identification results with respect to 
parametric assumptions. Structurally, it is assumed that 
the graph in Figure~\ref{fig:nonparamgraph} holds. 
Generally, identification of single-treatment quantities is not 
feasible in such settings. 
However, if researchers possess only factorial or observational 
distributions individually, identification of single-treatment probabilities as 
well as the ATE is possible only if they can justify that 
the real-world model is linear without interactions. Otherwise, 
this is achievable only when researchers have both 
factorial and observational distributions in the interactive linear case. 
Finally, if the assumption of no interaction holds, the ATE 
can be identified using both distributions, but single-treatment probabilities cannot.}
\label{table:idresults}
\begin{tabular}{ll|cccc}
\toprule
Data & Estimand & \makecell{General} & \makecell{AS (No Interaction)} & \makecell{Linearity \\ + Interaction} & \makecell{Linearity \\ + AS} \\
\midrule 
\multirow{2}{*}{Factorial} & $Y_a$ & \makecell{No (Thm.~\ref{theorem:nonidmonol}  ) } &  \makecell{No (Thm.~\ref{theorem:nointeractmono}  ) } & \makecell{No (Thm.~\ref{theorem:nointeractmono}  ) } & \makecell{Yes (Cor.~\ref{corollary:linearity})} \\
 & ATE & \makecell{No (Thm.~\ref{theorem:nonidmonol}  ) } & \makecell{Yes (Thm.~\ref{theorem:as}  ) } & \makecell{No (Thm.~\ref{theorem:nointeractonlyfact}  ) } & \makecell{Yes (Cor.~\ref{corollary:linearity})} \\
 \midrule
\multirow{2}{*}{Observ.}  & $Y_a$ & \makecell{No (Thm.~\ref{theorem:nonidmonol}  ) } & \makecell{No (Thm.~\ref{theorem:nointeractmono}  ) } & \makecell{No (Thm.~\ref{theorem:nointeractmono}  ) } & \makecell{No (Cor.~\ref{corollary:linearity})} \\
 & ATE & \makecell{No (Thm.~\ref{theorem:nonidmonol}  ) } & \makecell{No (Thm.~\ref{theorem:nointeractonlyfact}  ) } & \makecell{No (Thm.~\ref{theorem:nointeractonlyfact}  ) } & \makecell{No (Cor.~\ref{corollary:linearity})} \\
  \midrule
\multirow{2}{*}{Both} & $Y_a$ & \makecell{No (Thm.~\ref{theorem:nonidmonol}  ) } &  \makecell{No (Thm.~\ref{theorem:nointeractmono}  ) } & \makecell{Yes (Thm.~\ref{theorem:lininteract})} & \makecell{Yes (Cor.~\ref{corollary:linearity})} \\
 & ATE & \makecell{No (Thm.~\ref{theorem:nonidmonol}  ) } &  \makecell{Yes (Thm.~\ref{theorem:as}  ) } &  \makecell{Yes (Thm.~\ref{theorem:lininteract})} & \makecell{Yes (Cor.~\ref{corollary:linearity})} \\
\bottomrule
\end{tabular}
\end{table}

\subsection{Sufficient structural assumptions for point identifiability}

F-bias is typically problematic when one is not willing to make 
parametric assumptions.  However,
there are cases when making structural and/or functional assumptions might be justifiable by researchers. 
If one is willing to relax confounding of some variables in the single-treatment 
target population, then the problem is eliminated. In fact, 
the literature has already developed methods to deal with the problem of identifying causal quantities 
from multiple data sources with different interventional patterns \citep{bareinboim2016causal,lee2020general,lee2020identification}.
Here I illustrate three typical cases where single-treatment quantities are identifiable
given structural assumptions (non-confounding).

There are three typical cases where single-treatment effects are 
identifiable given structural assumptions (non-confounding) over the 
target population. 
 In Figure \ref{examplefbias0}-a,  
 for instance, $B$ can be confounded with $Y$, 
 but there is no effect from $B$ on $Y$. In cases represented by this graph, 
 then $\Pr(Y_{a, b} = y) = \Pr(Y_a = y)$.\footnote{This does not imply 
 that $\Pr(Y_{a} = y|B=b) = \Pr(Y_a = y)$. 
 Confusing causal interaction with effect modification is a conceptual blunder.} So,
  independently of the distribution of $B$ 
 employed to average $\Pr(Y_{a, b} = y)$, $\sum_{b \ \in \ \supp(B)} \Pr(Y_{a, b} = y) P(B = b) = \Pr(Y_a = y)$. 
 This example is relevant because it justifies adding new 
 attributes and interventions in experiments just to improve 
 realism. Without this justification, one might fear that 
 even irrelevant interventions might imply biased estimates, 
 but indeed they do not. For instance, a researcher might be concerned 
 that interventions such as whether a medication is prescribed as a 
 brand-name or a generic formulation could affect studies of that medication’s 
 impact on health outcomes. Because brand and generic formulations contain the 
 same active ingredient and are designed to be therapeutically equivalent, 
 such differences are often assumed not to causally influence the outcome 
 and are therefore ignored. However, in practice, brand versus 
 generic status may be correlated with patient characteristics, 
 insurance coverage, prescribing settings, or health-care access, 
 all of which are themselves associated with the outcome. As a result, 
 brand status can be confounded with the outcome even in the absence of a direct causal effect.

A second case is represented in Figure \ref{examplefbias0}-b.  
Here $B$ indeed causes $Y$, but they are unconfounded. 
In this case, then one can apply rule-2 of 
do-calculus \citep{pearl1995causal} to show that 
$\Pr(Y_{a, b} = y) = \Pr(Y_{a} = y|B = b)$, i.e. 
causal interaction is equivalent to effect modification. 
For this example, 
if we have the distribution of $B$ in the population, $\Pr(b)$, then $\Pr(Y_{a,b} = y) = \Pr(Y_a = y \ | \ B = b)$.  So, 
\begin{align*}
    \EX(Y_a) =  \sum_{b \ \in \ \supp(B)} \EX(Y_{a}|B = b)\Pr(B = b) = \sum_{b \ \in \ \supp(B)} \EX(Y_{a,b})\Pr(B = b),
\end{align*}
which is the same estimand in Equation~\ref{equation:pcme}, and the ATE is also identifiable. 
An example of this second scenario occurs when one is interested in evaluating 
the effect of a particular program, rather than a general factor. 
For instance, suppose one implements a randomized controlled trial (RCT) 
to evaluate the effect of providing cash transfers to 
students in a poor village, but simultaneously implements 
additional interventions related to an educational program on 
using the cash efficiently. In this case, the educational program 
would not typically be confounded in the real-world setting, 
which justifies the use of Equation \ref{equation:pcme}.

Finally, the third scenario is represented in Figure \ref{examplefbias0}-c 
and can be understood as a type of the DAG `b' case. 
Here, no variable is confounded with Y. Then the effect $\Pr(Y_a = y)$ 
can be identified by Equation~\ref{equation:uamce}, i.e. assuming 
a uniform distribution of $B$. Therefore, factorial data alone would be 
sufficient for identifying the single-treatment quantity.  
This third scenario corresponds to a case where the goal is to 
explore the effects of two programs without investigating the causal factors behind them. 
For instance, \citet{duflo2015education} tested the effect of two educational 
programs in Kenya on the probability of students dropping out of school, 
having an early marriage, or becoming pregnant. 
Those two programs are 1) educational subsidies, 
including free school uniforms, and 2) a national 
HIV curriculum to be taught to students. 
Because those programs are often interventional and cannot be confounded, 
then the single-treatment effect of each can be estimated without 
additional adjustment. However, if one intends to estimate 
not the effect of each program, but the effect 
of HIV education or certain subsidies in general, 
then researchers have to be more careful. 
The difference is subtle but is relevant for identifiability and estimation. 
This third situation is relevant in the scientific practice, because 
although the evaluation of a program might be successful, 
the ultimate goal of understanding the effect of a particular factor 
on a general population might not be achieved, especially in presence of 
confounding with other factors.

\begin{figure}
    \caption{Cases where $\Pr(Y_a)$ is point identifiable given $\Pr(Y_{a,b})$ and $\Pr(Y,A,B)$.}
        \label{examplefbias0}
    \centering
\begin{tikzpicture}
\begin{scope}[xshift=-7cm]
  \node[state]  (z) at (0,0) {$B$};
    \node[state]  (y) [below right =of z] {$Y$};
    \node[state]  (x) [below left =of z] {$A$};
    \path (x) edge (y);
    \path[bidirected] (x) edge[bend right=30] (y);
    \path[bidirected] (z) edge[bend left=30] (y);
    \end{scope}
        \node[anchor=center] at (-7, -3) {(a)};
    \begin{scope}[xshift=-2cm]
  \node[state]  (z) at (0,0) {$B$};
    \node[state]  (y) [below right =of z] {$Y$};
    \node[state]  (x) [below left =of z] {$A$};
    \path (x) edge (y);
    \path (z) edge (y);
    \path[bidirected] (x) edge[bend right=30] (y);
    \end{scope}
    \node[anchor=center] at (-2, -3) {(b)};
    \begin{scope}[xshift=3cm]
  \node[state]  (z) at (0,0) {$B$};
    \node[state]  (y) [below right =of z] {$Y$};
    \node[state]  (x) [below left =of z] {$A$};
    \path (x) edge (y);
    \path (z) edge (y);
    \end{scope}
    \node[anchor=center] at (3, -3) {(c)};
\end{tikzpicture}
\end{figure}
To summarize the results from this section, 
I gather all of them in Table~\ref{table:idresults} 
concerning parametric assumptions and Table~\ref{table:idresultsstr} 
regarding structural assumptions. Both tables serve as guides to 
assist researchers in determining which assumptions support particular conclusions.

\begin{table}[h]
    \centering
    \caption{Point-identification results of $\EX(Y_a)$ with respect to structural assumptions. Columns indicate which distributions are sufficient for identification (only observational, only factorial, or both). When necessary, cells indicate which rule of do-calculus enable identification, or if confounding prevents it. Nonparametrically (Fig.~\ref{fig:nonparamgraph}), neither factorial nor observational distributions allow for the identification of single-treatment results (no assumption). If one assumes that $B$ does not cause $Y$ (Fig.~\ref{examplefbias0}-a), then only the factorial distribution is adequate for identification, as per rule-3 of do-calculus. If one assumes that $B$ and $Y$ are not confounded (Fig.~\ref{examplefbias0}-b), then both factorial and observational data are necessary for identification. Finally, if neither $B$ nor $A$ are confounded with $Y$ (Fig.~\ref{examplefbias0}-c), then observational and factorial data become equivalent, leading to identification in both cases. }
\label{table:idresultsstr}
\begin{tabular}{ll|ccc}
\toprule
Assumption & Figure & \makecell{Observational} & \makecell{Factorial} & \makecell{Both}  \\
\midrule 
No assumption & Fig.~\ref{fig:nonparamgraph} & \makecell{No (Thm.~\ref{theorem:nonidmonol}}  )  &  \makecell{No (Thm.~\ref{theorem:nonidmonol}}  )  & \makecell{No (Thm.~\ref{theorem:nonidmonol}  ) }  
 \\
 \midrule
$B$ does not cause $Y$  & Fig.~\ref{examplefbias0}-a & \makecell{No (confounded) } & \makecell{ Yes (rule 3) } & \makecell{ Yes (rule 3) }    \\
  \midrule
Unconfounded $B$ and $Y$ & Fig.~\ref{examplefbias0}-b & \makecell{No (confounded) } &  \makecell{No (confounded) } & \makecell{Yes (rule 2)}  \\
 \midrule
Unconfounded $A$,$B$, and $Y$ & Fig.~\ref{examplefbias0}-c & \makecell{Yes (rule 1) } &  \makecell{Yes (rule 3) } & \makecell{Yes (rule 3)}  \\
\bottomrule
\end{tabular}
\end{table}

\section{Partial Identification}

\subsection{Closed-form bounds}

Up to this point, I have highlighted the challenges in 
identifying single-treatment quantities from factorial experiments. 
In particular, I showed that some assumptions are sufficient for 
point-identification, but they might be too restrictive in practice. 
If those assumptions are thought to fail, or if researchers are 
in doubt about their validity, 
then point-identification is not achievable. 

When these assumptions are not justifiable,
I propose two solutions: 
a) partial identification, 
and b) sensitivity analysis. 
This section focuses on the partial identification solution, 
deriving sharp bounds for the desired quantities. 
The next section will address the sensitivity analysis solution.

Partial identification, in general terms, refers to deriving a range of possible values for 
a quantity of interest. 
In this sense, it is considered a generalization of point 
identification, which is a case where the range 
consists of only one real number value. 
In other words, in a statistical model that imposes a 
number of constraints over probability distributions, 
one will collect only the distributions that satisfy those 
restrictions and compute the set of answers for a quantity of interest. 
Over this set of answers, one will calculate the best and 
worst-case scenarios in the form of sharp upper and lower bounds (maxima and minima).

In order to calculate the sharp bounds, I use the same procedure as 
the algorithm \emph{autobounds}, derived by \citet{duarte2023automated}. 
This approach consists in 
reducing causal questions with discrete data to polynomial programming 
problems, for which it is possible to sharply bound causal 
effects using efficient dual relaxation and spatial branch-and-bound techniques.  
From the perspective of an applied researcher, \emph{autobounds} 
is straightforward to use. The user (1) provides a causal diagram 
describing the process under study; (2) states assumptions; 
(3) states the quantity of interest; and 
(4) inputs available data, however incomplete or mismeasured. 
The algorithm then outputs the most precise possible answer 
given these parameters---which may be a range of indistinguishable, 
observationally equivalent possibilities. Thus, the 
method fully automates the process of computing \textit{causal bounds}---i.e., 
rigorously identifying all possible answers that cannot be ruled 
out by an observed dataset.  These results are made possible by 
the theoretical work proving that any causal problem can be transformed 
into an equivalent polynomial programming problem. Then, it becomes 
straightforward to derive a general algorithm for doing the problem 
reduction, which opens the door to a vast range of highly 
efficient polynomial programming solution techniques such as 
spatial branch-and-bound and linear programming relaxations. 
This method is an extension of the original \citet{balke1997bounds}'s 
linear programming technique and employs recent 
results in causality \citep{evans2018margins, rosset2018universal}.  

In particular, provided that the models considered here can 
be reduced to linear programs, I use symbolic solvers 
to provide closed-form solutions. For simplicity, assume only 
two binary treatments. More complex cases can be 
derived using \emph{autobounds} directly. 
The most general cases are illustrated in the next two 
theorems: the first one is applicable if one has only 
the factorial distribution, and the second one applies 
if one also has the observational distribution.

\begin{theorem}
Suppose $A, B, Y$ are binary random variables and one has only factorial data -- $\Pr(Y_{a,b})$. The sharp bounds for the expected value of $Y$ with intervention on $A$, $\Pr(Y_a)$, are:
\begin{align*}
\max\{0, 1 - \sum_b \Pr(Y_{ab} \neq y)\} \leq \Pr(Y_a) \leq \min\{\sum_b \Pr(Y_{ab} = y), 1\}
\end{align*}

and the bounds for the ATE ( $\EX(Y_{a_1} - Y_{a_0})$  ) are:
\begin{align*}
\max \begin{cases}
-1 \\
 - \EX(Y_{a_0b_0}  - Y_{a_0b_1}) \\
 \EX(Y_{a_1b_0}  + Y_{a_1 b_1}) - 2 \\
  \EX(Y_{a_1b_0} + Y_{a_1b_1} - Y_{a_0b_0} - Y_{a_0b_1} ) - 1
    \end{cases}  \leq 
    \text{ATE} \leq 
    \min 
    \begin{cases}
    1 \\
 2 -    \EX(Y_{a_0b_1}  + Y_{a_0b_0}) \\
   \EX(Y_{a_1b_0} + Y_{a_1b_1} )  \\ 
   1 + \EX(Y_{a_1b_0} + Y_{a_1b_1} - Y_{a_0b_0} - Y_{a_0b_1})
    \end{cases}
\end{align*}

\end{theorem}

\begin{theorem}
Suppose $A, B, Y$ are binary random variables and we have both factorial -- $\Pr(Y_{a,b})$ and observational data -- $\Pr(A,B,Y)$. Let $x' = 1 - x$ for any $x$. Then the sharp bounds for a quantity such as the expected value of $Y$ given intervention on $A$, $E(Y_a)$ are:
\begin{align*} \max  
\begin{cases} 0, \\
\Pr(Y_{ab'} = y) - \Pr(a',b) - \Pr(a,b,y') , \\
\Pr(a,y) , \\
\Pr(Y_{ab} = y') - \Pr(b',y') -  \Pr(a',b',y), \\
\Pr(Y_{ab'} = y) - \Pr(Y_{ab} = y')
\end{cases} \leq \EX(Y_a)  \leq \min
\begin{cases} 1, \\ 
\Pr(a',b,y') + \Pr(b,y) + \Pr(Y_{ab'} = y), \\
1 - \Pr(a,y'), \\
\Pr(Y_{ab'} = y) + \Pr(Y_{ab} = y), \\
\Pr(a',b') + \Pr(a,b',y) + \Pr(Y_{ab} = y)
\end{cases}
\end{align*}

and the sharp lower and upper bounds for the ATE are respectively:
\begin{align*}
\max \begin{cases}
- \Pr(a_1,y_0) - \Pr(a_0, y_1)  \\
\Pr(a_1,y_1) - \Pr(Y_{a_0b_0} = y_1) - \Pr(Y_{a_0b_1} = y_1)  \\
\Pr(a_1,b_1) - \Pr(a_1,y_0) - \Pr(a_0,b_0, y_1) - \Pr(Y_{a_0b_1} = y_1) \\
\Pr(a_1,b_0) - \Pr(a_1,y_0) - \Pr(a_0,b_1, y_1) - \Pr(Y_{a_0b_0} = y_1) \\
\Pr(a_0, y_0) - \Pr(Y_{a_1b_0} = y_0) - \Pr(Y_{a_1b_1} = y_0) \\
 \Pr(Y_{a_1b_0} = y_1) - \Pr(Y_{a_1b_1} = y_0) - \Pr(Y_{a_0b_0} = y_1) - \Pr(Y_{a_0b_1} = y_1) \\
  \Pr(Y_{a_1b_0} = y_1) - \Pr(Y_{a_1b_1} = y_0) - \Pr(Y_{a_0b_1} = y_1) - \Pr(a_0, b_0, y_1) - \Pr(a_1, b_0) \\
    \Pr(Y_{a_1b_0} = y_1) - \Pr(Y_{a_1b_1} = y_0) - \Pr(Y_{a_0b_0} = y_1) - \Pr(a_0, b_1, y_1) - \Pr(a_1, b_1) \\
    \Pr(a_0, b_1) - \Pr(a_1, b_0, y_0) - \Pr(a_0, y_1) - \Pr(Y_{a_1b_1} = y_0) \\ 
    \Pr(Y_{a_1b_1} = y_1) - \Pr(Y_{a_0b_0} = y_1) - \Pr(Y_{a_0b_1} = y_1) - \Pr(a_1, b_0, y_0) - \Pr(a_0, b_0) \\
\Pr(Y_{a_1b_1} = y_1) - \Pr(Y_{a_0b_1} = y_1)  - \Pr(b_0) - \Pr(a_1, b_0, y_0) - \Pr(a_0, b_0, y_1) \\
\Pr(a_1, b_0, y_1) + \Pr(a_0, b_1, y_0) - \Pr(Y_{a_1b_1} = y_0) - \Pr(Y_{a_0b_0} = y_1) \\
\Pr(a_0, b_0) - \Pr(a_1, b_1, y_0) - \Pr(a_0, y_1) -  \Pr(Y_{a_1b_0} = y_0) \\
\Pr(Y_{a_1b_0}=y_1) - \Pr(Y_{a_0b_0}=y_1) - \Pr(Y_{a_0b_1}=y_1) - \Pr(a_0,b_1) - \Pr(a_1,b_1,y_0) \\
\Pr(a_1,b_1,y_1) + \Pr(a_0,b_0,y_0) - \Pr(Y_{a_1b_0}=y_0) - \Pr(Y_{a_0b_1}=y_1) \\
\Pr(Y_{a_1b_0}=y_1) - \Pr(Y_{a_0b_1}=y_1) - \Pr(a_0,b_1) - \Pr(a_1,b_1,y_0) - \Pr(a_0,b_1,y_1) - \Pr(a_1,b_1) 
    \end{cases} 
    \end{align*}
    and \begin{align*} 
    \min  \begin{cases}
        \Pr(a_1,y_1) + \Pr(a_0, y_0) \\ 
        \Pr(Y_{a_0b_0}=y_0) + \Pr(Y_{a_0b_1}=y_0) - \Pr(a_1,y_0) \\
        \Pr(Y_{a_0b_1}=y_0)  + \Pr(a_0,b_0,y_0) + \Pr(a_1,b_0) - \Pr(a_1,y_0)\\
        \Pr(Y_{a_0b_0}=y_0) + \Pr(a_0,b_1,y_0) + \Pr(a_1,b_1) - \Pr(a_1,y_0) \\
        \Pr(Y_{a_1b_0}=y_1) + \Pr(Y_{a_1b_1}=y_1) - \Pr(a_0,y_1) \\
        \Pr(Y_{a_1b_0}=y_1) + \Pr(Y_{a_1b_1}=y_1) + \Pr(Y_{a_0b_0}=y_0) - \Pr(Y_{a_0b_1}=y_1)
\\
        \Pr(Y_{a_1b_0}=y_1) + \Pr(Y_{a_1b_1}=y_1) - \Pr(Y_{a_0b_1}=y_1) + \Pr(b_0,y_0) + \Pr(a_1,b_0,y_1)    \\
        \Pr(Y_{a_1b_0}=y_1) + \Pr(Y_{a_1b_1}=y_1) - \Pr(Y_{a_0b_0}=y_1) + \Pr(b_1,y_0) + \Pr(a_1,b_1,y_1)    \\
        \Pr(Y_{a_1b_1}=y_1) + \Pr(a_1,b_0,y_1) + \Pr(a_0,b_0,y_0) - \Pr(a_0,b_1, y_1) \\
        \Pr(Y_{a_1b_1}=y_1) + \Pr(Y_{a_0b_1}=y_0) - \Pr(Y_{a_0b_0}=y_1) + \Pr(a_0,b_0) + \Pr(a_1,b_0,y_1) \\
        \Pr(Y_{a_1b_1}=y_1) - \Pr(Y_{a_0b_1}=y_1) + \Pr(a_0,b_0) +  \Pr(b_0,y_0) + 2 \Pr(a_1,b_0,y_1)  \\
        \Pr(Y_{a_1b_1}=y_1) + \Pr(Y_{a_0b_0}=y_0) -  \Pr(a_1,b_0,y_0) - \Pr(a_0,b_1,y_1)
    \\
        \Pr(Y_{a_1b_0}=y_1) -  \Pr(a_0,b_0) + \Pr(a_1,b_1,y_1) + \Pr(a_0,y_0) \\ 
        \Pr(Y_{a_1b_0}=y_1) - \Pr(Y_{a_0b_0}=y_1) + \Pr(Y_{a_0b_1}=y_0) + \Pr(a_0,b_1) +   \Pr(a_1,b_1,y_1) \\ 
        \Pr(Y_{a_1b_0}=y_1) - \Pr(Y_{a_0b_1}=y_1) +  \Pr(a_0,b_1) + \Pr(a_1,y_1) + \Pr(b_0,y_0) \\ 
        \Pr(Y_{a_1b_0}=y_1) - \Pr(Y_{a_0b_0}=y_1) +  \Pr(a_0,b_1) +  \Pr(b_1,y_0) + 2 \Pr(a_1,b_1, y_1)
    \end{cases}
    \end{align*}
\end{theorem}

In the appendix, I have also derived bounds for estimands 
such as $\EX(Y_a)$ and the ATE in two other scenarios: 
one with only factorial data, and another with both 
observational and factorial data, under the monotonicity assumption.
There are notable limitations to those 
scenarios for cases involving treatments with more than two 
levels and other causal structures. However, these 
limitations can be addressed by employing the 
provided code, which makes use of \emph{autobounds}.

Despite being bounded in scenarios with binary outcomes, 
these bounds can be extended to the discrete case 
efficiently. One simply needs to compute bounds for each 
possible outcome of $y$. For instance, if $Y$ has 
three levels $y_0, y_1, y_2$, one calculates the 
bounds three times: first for $\Pr(Y_{a} = y_0)$, 
then for $\Pr(Y_{a} = y_1)$, and $\Pr(Y_{a} = y_2)$. 
During each bound computation, the other two levels are 
aggregated against the main one. When the outcome is 
continuous, a useful strategy suggested by both \citet{balke1997bounds} 
and \citet{levis2023covariate} is to calculate causal bounds 
for $\Pr(Y_a < k)$, for all $k$ within the bounded range of $Y$. 
In practice, this approach implies some form of discretization, 
ensuring valid bounds even if they are no longer sharp.

\subsection{Sensitivity Analysis}

In section 2, I demonstrated that assuming no 
interaction enables us to identify the ATE. 
However, justifying this assumption might pose challenges. 
This necessitates procedures to assess the extent to which our 
results hinge on this assumption. To address this, I propose a sensitivity analysis method. 

The concept behind this sensitivity analysis acknowledges the strong 
nature of the AS assumption. 
Instead of outright accepting this, one could restrict the 
proportion of units exhibiting some level of interaction. 
For example, a parameter like $\theta = 0.2$ acknowledges 
the potential presence of interactive units while also indicating 
that they may not exceed 20\% of the population. 
Additionally, one can determine the maximum proportion of interactive 
units that could nullify the ATE or push it towards a specific value. 
Therefore, this procedure can be applied when presenting 
results on single-treatment effects, particularly with factorial data.

\begin{figure}
    \centering
    \caption{Sensitivity analysis of the ATE, constrained by the proportion of non-interactive units, is presented.  The details of the model are introduced in the Supplementary Material (Example 2). The dashed line represents the actual ATE of the model (0.58), while the black lines depict strict bounds corresponding to varying constraints on the maximum proportion of non-interactive units (x-axis). A maximum constraint of 0 implies no interaction, resulting in exact point identification of the ATE. As this maximum constraint increases, the bounds extend to the maximum limits reported in Theorems~\ref{theorem:boundsmain} and~\ref{theorem:boundsonlyfact}.}
    \includegraphics[scale=0.7]{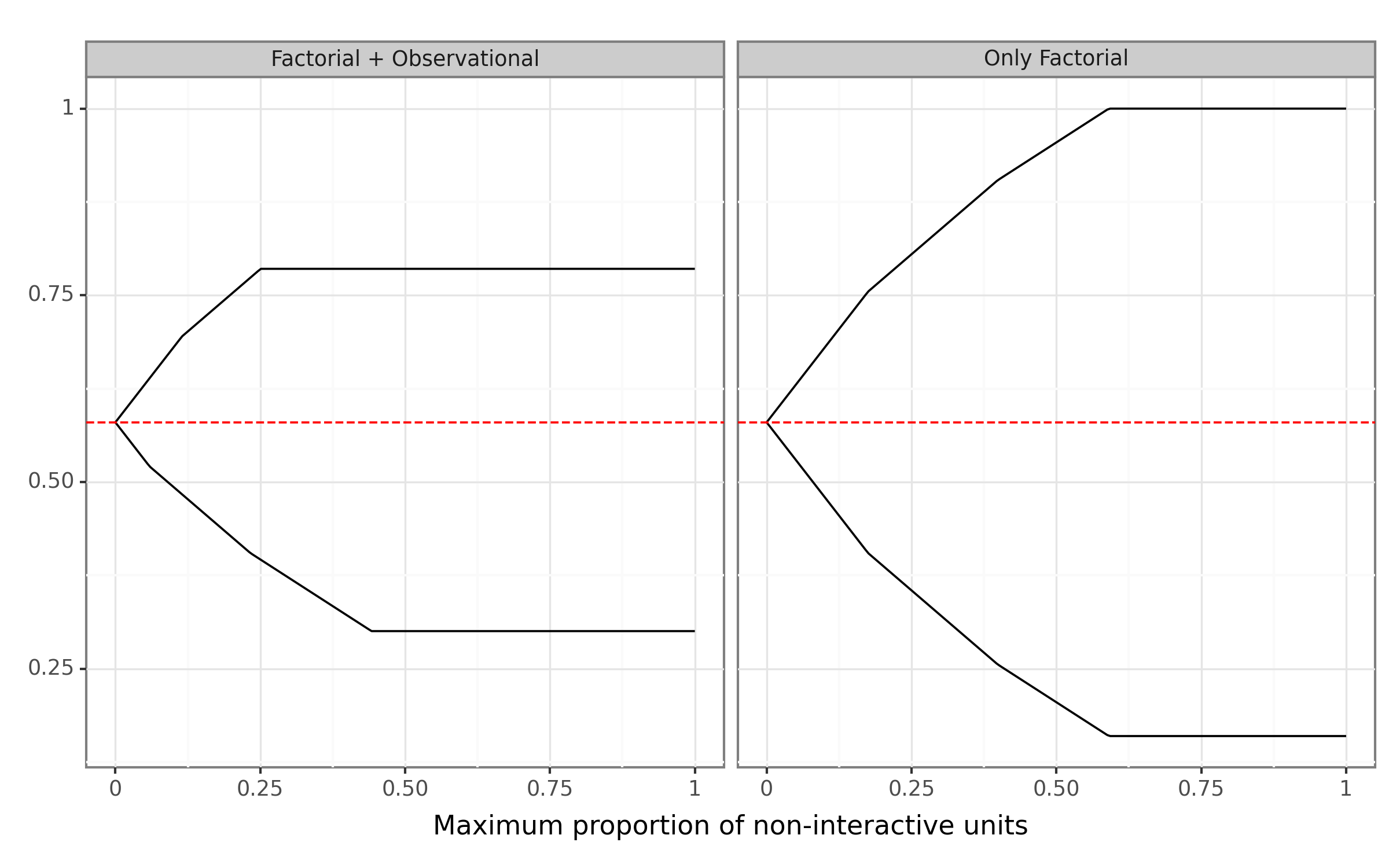}
        \label{figsensitivity}
\end{figure}

Similar to the preceding procedure, the objective of this analysis 
is to determine precise bounds for the ATE. 
However, in this scenario, the strata associated with interaction 
will be restricted to a maximum of $\theta$, 
while the rest are considered freely. Using the 
abbreviation $y_{ijkl}$ for the 
strata $Y_{a_0,b_0} = i, Y_{a_0,b_1} = j, Y_{a_1,b_0} = k, Y_{a_1,b_1} = l$, 
we have six non-interactive strata: $y_{0000}, y_{0011}, y_{0101}, y_{1010}, y_{1100}, y_{1111}$, 
with all others being interactive. 
The nature of this problem leads to a linear program with 
non-zero inequalities, making analytical solution more challenging. 
Hence, the analysis presented in the simulation section will 
rely on the computational capabilities of the autobounds package. 
A possible application is demonstrated in Figure~\ref{figsensitivity} and discussed below.

\section{Simulation}
To illustrate the proposed method, 
I simulate two examples described in the Supplementary Material. 
The first example involves a structural model with no explicit constraints, 
while the second one is generated using probabilities 
over principal strata \citep{greenland1986identifiability, frangakis2002principal} to 
enforce both monotonicity and no interaction. Then I generate firstly for each 
model an example dataset with size equal to $1,000$, 
both for the observational and for the factorial case (total of $2,000$).

Previously, I discussed identification results tailored to 
scenarios where researchers have access to population quantities. 
Here, I discuss the challenge of estimating bounds when dealing with finite samples. 
In such cases, an additional hurdle arises as it's not straightforward 
to estimate confidence intervals for extremas of distributions. Firstly,
the main issue is that the bounds are not smooth functions of the
underlying distributions, which complicates the application of standard
asymptotic theory. It is known that if one maximizes (or minimizes) over 
many candidate functions that are smooth, the resulting function may not be smooth. 
This eschews the use of both delta method and bootstrap. Secondly, in practice, 
it is often assumed that the if one knows which candidate is the correct one in 
the population of interest or at least 
approximately knows it, then one can use standard asymptotic theory. 
However, in finite samples, the selected candidate may differ from the
population one with non-negligible probability, leading to incorrect inference. 
This issue has been explored extensively in the econometrics literature,
notably by \citet{andrews2010inference, 
bugni2010bootstrap, canay2010inference, chernozhukov2013intersection}.

Several solutions have been proposed to address this challenge.
For simplicity, I employ Bayesian Dirichlet posteriors to estimate the 
underlying distributions \citep{richardson2011transparent, marden2018implementation}. 
Specifically, I assume a Dirichlet prior with 
parameters equal to 1 for each cell of the joint distribution. 
After observing the data, I compute the posterior distribution,
which is also Dirichlet, with parameters equal to the observed counts plus 1.
Differently from bootstrap, which resamples the data from a multinomial distribution 
with parameters equal to the empirical frequencies,
the Dirichlet posterior can be interpreted as sampling from a 
multinomial distribution with parameters equal to the smoothed frequencies.
Despite not having frequentist guarantees, this approach is valid from a Bayesian perspective.

From the Dirichlet posterior distribution, I sample $1000$ times the underlying 
distributions and compute the bounds for each sample.
So I calculate 95\% confidence bounds 
both for the lower bounds and for the upper bounds. 

\begin{figure}
    \centering
    \caption{Estimated bounds using simulated data from examples 1 and 2. 
    The blue bars represent the estimated bounds, while the red lines indicate the 95\% credible region 
    for the bounds calculated using Dirichlet posteriors. 
    The dotted black line shows the true value of the ATE 
    for each example: 0.24 for the first and 0.58 for the second. 
    Example 2 explicitly contains only no interactive and monotonic units. 
    Therefore, the analysis for the first model only shows nonparametric bounds. 
    As expected, both bounds cover the real ATE, but those incorporating 
    both factorial and observational data are sharper than those with factorial data alone. 
    For the second example, we estimated three versions: 
    one assuming monotonicity, another assuming a maximum proportion of 0.15 non-interactive 
    units, and a nonparametric version. As expected, the 
    bounds in the nonparametric version are lower than in the other two. 
    In other words, justified restrictive assumptions lead to a quantitative gain in causal information. }
\includegraphics[scale=0.7]{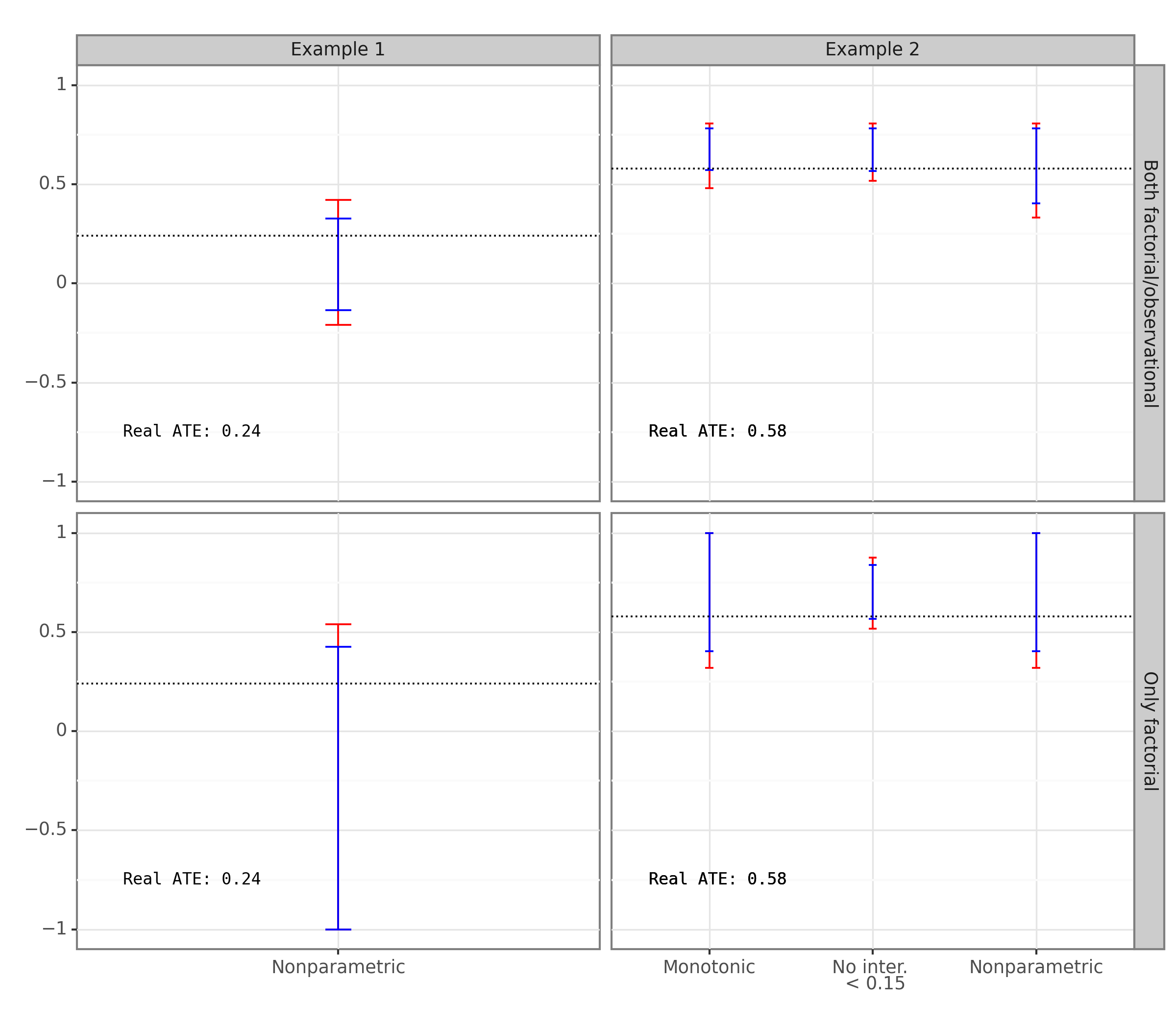}
    \label{fig:estimationresult}
\end{figure}

Example 1 was simulated nonparametrically using structural equations, 
without assuming non-interactivity and monotonicity. 
The model produced an ATE of 0.24. Two sets of results 
were estimated: firstly, nonparametric bounds assuming only factorial data, 
and secondly, nonparametric bounds using both factorial and observational data. 
These results are depicted in the left panel of Figure~\ref{fig:estimationresult}. 
The estimated bounds (blue lines) encompass the true ATE value (black lines). 
The version incorporating both distributions yields bounds superior to the second version. 
Additionally, I estimated a 95\% confidence region (in red) using resampling techniques. 
In this particular example, one cannot rule out that the real ATE is different from 0.

In example 2, I explicitly 
made assumptions of monotonicity and no interaction, 
with a true ATE of 0.58. 
This justifies estimating bounds based on these assumptions. 
As demonstrated, assuming no interaction can lead to point identification results. 
Rather than adopting this approach, I considered a scenario where one 
might wish to examine the sensitivity of the no-interaction assumption. 
In Figure~\ref{figsensitivity}, I illustrate bounds 
conditioned on the maximum proportion of non-interactive units permitted in the model. 
For instance, if no constraint is imposed and full interactions are 
allowed, then nonparametric bounds are obtained. 
All other cases lie between the full interaction and non-interactive point identifiable scenarios. 
After analyzing the graph, suppose one believes that a 
maximum of 15\% of non-interactive units is reasonable; in that case, the bounds can be evaluated accordingly, 
along with credible intervals. This result is depicted in the 
right panel of Figure~\ref{fig:estimationresult}. Additionally, 
bounds for the case assuming only monotonicity were included, 
showing marked improvements compared to the nonparametric scenario. 
However, all these cases, even when utilizing only the factorial distribution, 
provide an opportunity to estimate bounds greater than 0 (which can be signed) and encompass the real ATE.

\section{Discussion}

The paper's findings highlight that while factorial experiments 
can significantly reduce the costs associated 
with estimating multiple treatment effects, 
they also introduce complexities when extrapolating to 
 single-treatment scenarios where the other treatments are not manipulated. 
 Specifically, when considering interactions, 
 these effects become non-recoverable, without imposing strong assumptions. 

 Recognizing the challenges in justifying such assumptions,
 and a practical approach is to estimate sharp bounds. 
 To address this, I provided bounds for both the nonparametric case 
 and the case assuming monotonicity, 
 aiming to assist researchers in estimating effects under these conditions. 
 Moreover, I outlined a procedure for sensitivity analysis, 
 allowing researchers to explore other assumptions, 
 such as non-interaction, or to test the sensitivity 
 of results to a maximum proportion of non-interactive units. 
 For instance, by limiting interactive units to at most 15\%, 
 researchers can still derive meaningful bounds.
 That shows how useful information on external validity can be 
 extracted from factorial experiments.

\bibliography{refs}

@article{hainmueller2014causal,
  title={Causal inference in conjoint analysis: Understanding multidimensional choices via stated preference experiments},
  author={Hainmueller, Jens and Hopkins, Daniel J and Yamamoto, Teppei},
  journal={Political analysis},
  volume={22},
  number={1},
  pages={1--30},
  year={2014},
  publisher={Cambridge University Press}
}

@article{fisher1949design,
  title={The design of experiments},
  author={Fisher, Ronald A},
  year={1949},
  publisher={Oliver \& Boyd}
}

@article{andrews2010inference,
  title={Inference for parameters defined by moment inequalities using generalized moment selection},
  author={Andrews, Donald WK and Soares, Gustavo},
  journal={Econometrica},
  volume={78},
  number={1},
  pages={119--157},
  year={2010},
  publisher={Wiley Online Library}
}

@article{chernozhukov2013intersection,
  title={Intersection bounds: Estimation and inference},
  author={Chernozhukov, Victor and Lee, Sokbae and Rosen, Adam M},
  journal={Econometrica},
  volume={81},
  number={2},
  pages={667--737},
  year={2013},
  publisher={Wiley Online Library}
}

@article{bugni2010bootstrap,
  title={Bootstrap inference in partially identified models defined by moment inequalities: Coverage of the identified set},
  author={Bugni, Federico A},
  journal={Econometrica},
  volume={78},
  number={2},
  pages={735--753},
  year={2010},
  publisher={Wiley Online Library}
}

@article{canay2010inference,
  title={EL inference for partially identified models: Large deviations optimality and bootstrap validity},
  author={Canay, Ivan A},
  journal={Journal of Econometrics},
  volume={156},
  number={2},
  pages={408--425},
  year={2010},
  publisher={Elsevier}
}

@article{dasgupta2015causal,
  title={Causal inference from 2K factorial designs by using potential outcomes},
  author={Dasgupta, Tirthankar and Pillai, Natesh S and Rubin, Donald B},
  journal={Journal of the Royal Statistical Society Series B: Statistical Methodology},
  volume={77},
  number={4},
  pages={727--753},
  year={2015},
  publisher={Oxford University Press}
}

@article{greenland1986identifiability,
  title={Identifiability, exchangeability, and epidemiological confounding},
  author={Greenland, Sander and Robins, James M},
  journal={International journal of epidemiology},
  volume={15},
  number={3},
  pages={413--419},
  year={1986},
  publisher={Oxford University Press}
}

@article{lee2020identification,
  title={Identification methods with arbitrary interventional distributions as inputs},
  author={Lee, Jaron JR and Shpitser, Ilya},
  journal={arXiv preprint arXiv:2004.01157},
  year={2020}
}

@inproceedings{lee2020general,
  title={General identifiability with arbitrary surrogate experiments},
  author={Lee, Sanghack and Correa, Juan D and Bareinboim, Elias},
  booktitle={Uncertainty in artificial intelligence},
  pages={389--398},
  year={2020},
  organization={PMLR}
}

@article{blackwell2023noncompliance,
  title={Noncompliance and instrumental variables for 2 K factorial experiments},
  author={Blackwell, Matthew and Pashley, Nicole E},
  journal={Journal of the American Statistical Association},
  volume={118},
  number={542},
  pages={1102--1114},
  year={2023},
  publisher={Taylor \& Francis}
}

@article{frangakis2002principal,
  title={Principal stratification in causal inference},
  author={Frangakis, Constantine E and Rubin, Donald B},
  journal={Biometrics},
  volume={58},
  number={1},
  pages={21--29},
  year={2002},
  publisher={Oxford University Press}
}

@article{pashley2023causal,
  title={Causal inference for multiple treatments using fractional factorial designs},
  author={Pashley, Nicole E and Bind, Marie-Ab{\`e}le C},
  journal={Canadian Journal of Statistics},
  volume={51},
  number={2},
  pages={444--468},
  year={2023},
  publisher={Wiley Online Library}
}

@article{egami2018causal,
  title={Causal interaction in factorial experiments: Application to conjoint analysis},
  author={Egami, Naoki and Imai, Kosuke},
  journal={Journal of the American Statistical Association},
  year={2018},
  publisher={Taylor \& Francis}
}

@article{zhao2022regression,
  title={Regression-based causal inference with factorial experiments: estimands, model specifications and design-based properties},
  author={Zhao, Anqi and Ding, Peng},
  journal={Biometrika},
  volume={109},
  number={3},
  pages={799--815},
  year={2022},
  publisher={Oxford University Press}
}

@book{imbens2015causal,
  title={Causal inference in statistics, social, and biomedical sciences},
  author={Imbens, Guido W and Rubin, Donald B},
  year={2015},
  publisher={Cambridge university press}
}

@article{levis2023covariate,
  title={Covariate-assisted bounds on causal effects with instrumental variables},
  author={Levis, Alexander W and Bonvini, Matteo and Zeng, Zhenghao and Keele, Luke and Kennedy, Edward H},
  journal={arXiv preprint arXiv:2301.12106},
  year={2023}
}

@article{rosset2018universal,
  title={Universal bound on the cardinality of local hidden variables in networks},
  author={Rosset, Denis and Gisin, Nicolas and Wolfe, Elie},
  journal={Quantum Information \& Computation},
  volume={18},
  number={11-12},
  pages={910--926},
  year={2018},
  publisher={Rinton Press, Incorporated Paramus, NJ}
}

@article{evans2018margins,
  title={Margins of discrete Bayesian networks},
  author={Evans, Robin J},
  journal={The Annals of Statistics},
  volume={46},
  number={6A},
  pages={2623--2656},
  year={2018},
  publisher={Institute of Mathematical Statistics}
}

@article{de2022improving,
  title={Improving the external validity of conjoint analysis: The essential role of profile distribution},
  author={De la Cuesta, Brandon and Egami, Naoki and Imai, Kosuke},
  journal={Political Analysis},
  volume={30},
  number={1},
  pages={19--45},
  year={2022},
  publisher={Cambridge University Press}
}

@article{duflo2015education,
  title={Education, HIV, and early fertility: Experimental evidence from Kenya},
  author={Duflo, Esther and Dupas, Pascaline and Kremer, Michael},
  journal={American Economic Review},
  volume={105},
  number={9},
  pages={2757--97},
  year={2015}
}

@article{duarte2023automated,
  title={An automated approach to causal inference in discrete settings},
  author={Duarte, Guilherme and Finkelstein, Noam and Knox, Dean and Mummolo, Jonathan and Shpitser, Ilya},
  journal={Journal of the American Statistical Association},
  number={just-accepted},
  pages={1--25},
  year={2023},
  publisher={Taylor \& Francis}
}

@article{duflo2007using,
  title={Using randomization in development economics research: A toolkit},
  author={Duflo, Esther and Glennerster, Rachel and Kremer, Michael},
  journal={Handbook of development economics},
  volume={4},
  pages={3895--3962},
  year={2007},
  publisher={Elsevier}
}

@article{balke1997bounds,
  title={Bounds on treatment effects from studies with imperfect compliance},
  author={Balke, Alexander and Pearl, Judea},
  journal={Journal of the American Statistical Association},
  volume={92},
  number={439},
  pages={1171--1176},
  year={1997},
  publisher={Taylor \& Francis}
}

@article{bareinboim2016causal,
  title={Causal inference and the data-fusion problem},
  author={Bareinboim, Elias and Pearl, Judea},
  journal={Proceedings of the National Academy of Sciences},
  volume={113},
  number={27},
  pages={7345--7352},
  year={2016},
  publisher={National Acad Sciences}
}

@article{marden2018implementation,
  title={Implementation of instrumental variable bounds for data missing not at random},
  author={Marden, Jessica R and Wang, Linbo and Tchetgen, Eric J Tchetgen and Walter, Stefan and Glymour, M Maria and Wirth, Kathleen E},
  journal={Epidemiology},
  volume={29},
  number={3},
  pages={364--368},
  year={2018},
  publisher={LWW}
}

@article{richardson2011transparent,
  title={Transparent parameterizations of models for potential outcomes},
  author={Richardson, Thomas S and Evans, Robin J and Robins, James M},
  journal={Bayesian statistics},
  volume={9},
  pages={569--610},
  year={2011},
  publisher={Oxford University Press New York}
}

@article{pearl1995causal,
  title={Causal diagrams for empirical research},
  author={Pearl, Judea},
  journal={Biometrika},
  volume={82},
  number={4},
  pages={669--688},
  year={1995},
  publisher={Oxford University Press}
}
\newpage

\section*{Supplementary material}
\label{SM}

\vspace*{10pt}

\appendix


\section{Hypothetical Examples}

\subsection{Example 1}
\label{example:example1}

Suppose there are unobserved and observed variables. Assume that the unobserved -- represented by variables starting with $U$ -- are distributed by binomial distributions with the following parameters:  
\begin{align*}
    U_{AY} \sim Binom(0.65)\\
    U_{BY} \sim Binom(0.8)\\
    U_{Y_1} \sim Binom(0.95)\\
    U_{Y_2} \sim Binom(0.9)
\end{align*}
At last, the functions representing the observed variables $A, B,$ and $Y$ can be defined using logic gates, as they are all binary:
\begin{align*}
    A = U_{AY} \\
    B = U_{BY} \\
    Y =  U_{AY} \land \lnot( (U_{Y_1} \land B ) \oplus (U_{BY} \land U_{Y_1} )  ) \oplus ( A \oplus U_{Y_2} )\\
\end{align*}

Here, $\land$ denotes the conjunctive logic gate (AND), $\oplus$ denotes the exclusive disjunction (XOR), and $\lnot$ denotes the negation (NOT).

From this example, one can calculate quantities such as the real ATE, which is equal to $0.24$. However, the factorial experiment yields quantities tied to interventions on both $A$ and $B$ at the same time. In particular, the difference of the expected value of $Y$ given interventions on $A = a_1$ and $A = a_0$, given that we also had an intervention on $B = b$ are respectively, $-0.55$ and $0.042$, for the cases where $B = b_0$ and $B = b_1$. Using the estimator described by the first researcher -- which averages those differences according to a uniform distribution, i.e. $P(B=b_1)=P(B=b_0) = 0.5$ -- yields an estimate of $-0.25$. Alternatively, using the estimator described by the second researcher -- which employs the population distribution of $B$ -- yields an estimate of $-0.07$. In this example, both estimates are negative despite the positive ATE. There is no distribution over $B$ that combines the factorial differences into the real ATE in this setting. This suggests that attempts to use factorial experiments to answer queries concerning marginal treatments can be biased. I refer to the bias induced when extrapolating from \underline{f}actorial experiments to in-the-wild effects of isolated treatments as \emph{F-bias}.  

\subsection*{Example 2}
\label{example:example2}

Differently from the previous example, I simulate this case according to the specified canonical model and principal strata. The values of each parameter are available in table~\ref{tab:example2}.  I assign values for each strata of $A$ and $B$, represented in the columns, and to $Y$, represented in the rows. For notation economy, the strata for $Y$, $\Pr(Y_{a_0,b_0} 
= i, Y_{a_0,b_1} = j, Y_{a_1,b_0} = k,Y_{a_1,b_1} = l)$,  is abbreviated to $y_{ijkl}$. The cells indicate $\Pr(y_{ijkl} | a, b)$ for all values of $A$ and $B$. So, to obtain the joint probability, one has to multiply them by $\Pr(a,b)$, distributed as $\Pr(A=a_0, B=b_0) = 0.15, \Pr(A=a_0, B=b_1) = 0.5, \Pr(A=a_1, B=b_0) = 0.1, \Pr(A=a_1, B=b_1) = 0.25$.

In particular, all non-monotonic and non-interactive strata were set to have probability equal to 
$0$.  So only probability over the strata $y_{0000}, y_{0101}, y_{0011}, y_{1111}$ were allowed. 
This model yields an ATE $0.58$. Eventual calculations are provided in the attached python code.

\begin{table}
\caption{Strata for a causal model of $A$ and $B$ causing $Y$ and unobserved confounding. Cells indicate $\Pr(Y_{a_0,b_0} 
= i, Y_{a_0,b_1} = j, Y_{a_1,b_0} = k,Y_{a_1,b_1} = l| a, b)$ abbreviated to $\Pr(y_{ijkl}|a, b)$, where $y_{ijkl}$ is indicated by the rows and $a, b$ by each column.}
\begin{center}
\begin{tabular}{l|r|r|r|r}
 &  $a_{0}, b_0$                   & $a_0, b_1$    & $a_1, b_0$                   & $a_1, b_1$     \\
\hline
$y_{0000}$ & 0.05               & 0.09  & 0.05                & 0.12  \\
\hline
$y_{0001}$ & 0.0               & 0.0 & 0.0                & 0.0  \\
\hline
$y_{0010}$ & 0.0                & 0.0 & 0.0                  & 0.0    \\
\hline
$y_{0011}$ & 0.7                & 0.6 & 0.5                & 0.5  \\
\hline
$y_{0100}$ & 0.0               & 0.0   & 0.0                & 0.0    \\
\hline
$y_{0101}$ & 0.1               & 0.1  & 0.02                & 0.03  \\
\hline
$y_{0110}$ & 0.0               & 0.0   & 0.0                  & 0.0    \\
\hline
$y_{0111}$ & 0.0               & 0.0 & 0.0                & 0.0   \\
\hline
$y_{1000}$ & 0.0                  & 0.0   & 0.0                  & 0.0  \\
\hline
$y_{1001}$ & 0.0                  & 0.0   & 0.0                  & 0.0 \\
\hline
$y_{1010}$ & 0.0                  & 0.0   & 0.0                  & 0.0    \\
\hline
$y_{1011}$ & 0.0                  & 0.0   & 0.0                & 0.0  \\
\hline
$y_{1100}$ & 0.0                  & 0.0   & 0.0                  & 0.0    \\
\hline
$y_{1101}$ & 0.0                  & 0.0 & 0.0                  & 0.0 \\
\hline
$y_{1110}$ & 0.0               & 0.0   & 0.0                  & 0.0    \\
\hline
$y_{1111}$ & 0.15               & 0.21 & 0.43                & 0.35 
\end{tabular}
\end{center}
\label{tab:example2}
\end{table}

\newpage

\section{Non-identifiability results}
\label{appendix:nonid}

\begin{theorem}
\label{theorem:nonidmonol}
    Consider treatments $A$ and $B$, along with an outcome $Y$. Assume that both $A$ and $B$ independently cause $Y$, without causing each other. Additionally, suppose the effects of both $A$ and $B$ on $Y$ are monotonic, meaning $\Pr(Y_{a_0}=y_1, Y_{a_1}=y_0) = \Pr(Y_{b_0}=y_1, Y_{b_1}=y_0) = 0$. Under these conditions, without imposing any other assumptions, such as absence of confounding between $A$, $B$, and $Y$, the probability distribution $\Pr(Y_a)$, representing the outcome under different levels of treatment $A$, is not identifiable solely from the joint distribution $\Pr(Y_{a, b})$ and $\Pr(Y, A, B)$. Furthermore, the ATE, defined as the difference in probabilities between different treatment levels, $\Pr(Y_a = y) - \Pr(Y_{a'} = y)$, remains unidentifiable. Finally, these quantities are not identifiable even if we relax the assumption of monotonicity or if we only have observational or factorial data alone.
\end{theorem}
\begin{proof}

The proof employs a counterexample using binary variables $A$, $B$, and $Y$. Two models, denoted $M_0$ and $M_1$, are constructed such that they exhibit monotonicity of $A$ and $B$ on $Y$ and share the same distributions over $\Pr(A,B,Y)$ and $\Pr(Y_{a,b})$. However, they diverge in terms of the distributions of $\Pr(Y_{a})$ and $\Pr(Y_{a}) - \Pr(Y_{a'})$.

\textbf{Model specification:}

Consider models $M_0$ and $M_1$, where errors in both models are determined by three independent variables: binary $U_{ay}$, binary $U_{by}$, and ternary $U_y$. The probabilities are specified as $\Pr(U_{ay} = 1) = \Pr(U_{by} = 1) = 0.5$, $\Pr(U_y = 2) = 1/5$, and $\Pr(U_y = 1) = \Pr(U_y = 0) = 2/5$.

In both models, the functions determining $A$ and $B$ are set equal to $A = U_{ay}$ and $B = U_{by}$, meaning both variables are equal to the respective errors.

Now, let the outcome $Y$ for model $0$, $Y_{M_0}$,  be determined by: 
     \begin{align*}
         Y_{M_0} = \begin{cases} 
         \max\{A, B\}, & \text{ if } U_{ay} = U_{by} = 0 \ \text{ and } \ U_{y} = 2, \\
          B, & \text{ if } U_{ay} = U_{by} = 0 \ \text{ and } \ U_{y} = 1, \\            1, & \text { otherwise }
         \end{cases}
     \end{align*}

Also let $Y$ for model $M_1$, $Y_{M_1}$ be equal to:

     \begin{align*}
         Y_{M_1} = \begin{cases} 
         \max\{A, B\}, & \text{ if } U_{ay} = U_{by} = 0 \ \text{ and } \ U_{y} \neq 0, \\
        \max\{1 - A, B\}, & \text{ if } U_{ay} = 0, U_{by} = 1 \ \text{ and } \ U_{y} \neq 0, \\
               1, & \text { otherwise }
         \end{cases}
     \end{align*}

\textbf{Observational Data:}

For data generated by model $M_0$, if $A$ or $B$ equals $1$, then $Y$ must also equal $1$, implying probabilities of $0$ for scenarios where $Y = 0$ under those conditions and $0.25$ for scenarios where $Y = 1$ under the same conditions. When $A = B = 0$, there are three possibilities: if $U_y = 2$ or $U_y = 1$, then $Y = 0$, and if $U_y = 0$, then $Y = 1$. This yields probabilities of $3/5$ for $Y = 0$ and $2/5$ for $Y = 1$.

For data generated by model $M_1$, if $A = 1$, then $Y$ must equal $1$, resulting in probabilities of $0$ for scenarios where $Y = 0$ and $0.25$ for scenarios where $Y = 1$. If $A = 0$ and $B = 1$, then $Y$ is always $1$, independently of $U_y$, leading to probabilities of $0$ for $Y = 0$ and $0.25$ for $Y = 1$. When $U_{ay} = U_{by} = 0$, there are two possibilities: if $U_y = 2$ or $U_y = 1$, then $Y = 0$, and if $U_y = 0$, then $Y = 1$, resulting in probabilities of $3/5$ for $Y = 0$ and $2/5$ for $Y = 1$.

\textbf{Factorial Data:}

In both models $M_0$ and $M_1$, when $B$ is forced to be $1$, $Y$ is always $1$. If $B$ is forced to be $0$, and $A$ is also forced to be $0$, then $Y$ is $0$ when $U_{ay} = U_{by} = 0$ and $U_y \neq 0$, resulting in probabilities of $0.15$ for $Y = 0$ and $0.85$ for $Y = 1$. If $A$ is forced to be $1$, then $Y$ is $0$ when $U_{ay} = U_{by} = 0$ and $U_y = 1$, leading to probabilities of $0.1$ for $Y = 0$ and $0.9$ for $Y = 1$.

\textbf{Monotonicity:}

Additionally, both models satisfy monotonicity with respect to $A$ and $B$. There are no cases where $Y_{a = 1} = 0$ and $Y_{a = 0} = 1$ in model $M_1$, and in model $M_0$, the only such case occurs when $Y = B$, which remains unchanged if $A$ is set to $0$. Similarly, there are no cases where $Y_{b = 1} = 0$ and $Y_{b = 0} = 1$ due to the nature of $Y$ in both models.

These analyses reveal that the observational and factorial quantities are equal in both monotonic models. Now we consider the single-treatment quantities. A summary of all those quantities is included in Table~\ref{tab:probmonotonicce}.

\textbf{Single-treatment quantities}:

When $A$ is set to $0$ regardless of $B$ in both models $M_0$ and $M_1$, only one scenario results in $Y = 0$: when $U_{ay} = U_{by} = 0$ and $U_y \neq 0$. This yields $\Pr_{M_0}(Y_{A=0}=0) = \Pr_{M_1}(Y_{A=0}=0) = 0.15$. However, when $A$ is forced to be $1$, in model $M_0$, $Y$ consistently equals $1$ across all scenarios. Yet, in model $M_1$, there's a case where $Y = 0$ when $U_{ay} = U_{by} = 0$ and $U_y = 1$, with a probability of $0.1$. Thus, $\Pr_{M_0}(Y_{A=1}=1) = 1$, different from $\Pr_{M_1}(Y_{A=1}=1) = 0.9$. Consequently, $ATE_{M_0} = 0.15$, in contrast to $ATE_{M_1} = 0.05$. This disparity demonstrates that both $\Pr(Y_{A=1})$ and the ATE lack point identifiability.

\begin{table}[htbp]
\centering
\caption{Calculated probabilities for models $M_0$ and $M_1$ as 
presented in the counterexample detailed in Theorem~\ref{theorem:nonidmonol}. 
Despite both models being monotonic, they agree on all observational and factorial 
(2-treatment) quantities. However, they diverge regarding $\Pr(Y_{A=1})$, 
making not only this probability but also the ATE non-identifiable}
\label{tab:probmonotonicce}
\begin{tabular}{@{}l|ccc@{}}
\toprule
\textbf{Probability} & \textbf{Model $M_0$} & \textbf{Model $M_1$} \\
\midrule
\multicolumn{3}{@{}l}{\textbf{Observational}} \\
\midrule
$\Pr(A = 0, B = 0, Y = 0)$ & 0.15 & 0.15 \\
$\Pr(A = 0, B = 1, Y = 0)$ & 0 & 0 \\
$\Pr(A = 1, B = 0, Y = 0)$ & 0 & 0 \\
$\Pr(A = 1, B = 1, Y = 0)$ & 0 & 0 \\
$\Pr(A = 0, B = 0, Y = 1)$ & 0.1 & 0.1 \\
$\Pr(A = 0, B = 1, Y = 1)$ & 0.25 & 0.25 \\
$\Pr(A = 1, B = 0, Y = 1)$ & 0.25 & 0.25 \\
$\Pr(A = 1, B = 1, Y = 1)$ & 0.25 & 0.25 \\
\midrule
\multicolumn{3}{@{}l}{\textbf{2-Treatment}} \\
\midrule
$\Pr(Y_{A=0, B = 0} = 1)$ & 0.85 & 0.85 \\
$\Pr(Y_{A=0, B = 1} = 1)$ & 1 & 1 \\
$\Pr(Y_{A=1, B = 0} = 1)$ & 0.9 & 0.9 \\
$\Pr(Y_{A=1, B = 1} = 1)$ & 1 & 1 \\
\midrule
\multicolumn{3}{@{}l}{\textbf{1-Treatment}} \\
\midrule
$\Pr(Y_{A=1}=1)$ & 1 & 0.9 \\
$\Pr(Y_{A=0}=1)$ & 0.85 & 0.85 \\
\bottomrule
\end{tabular}
\end{table}

The non-identification result demonstrated persists if we relax monotonicity or if we have only factorial, as those scenarios are less restrictive. 
\end{proof}

\begin{theorem}
    \label{theorem:nointeractmono}
  Consider treatments $A$ and $B$, along with an outcome $Y$. Assume that both $A$ and $B$ independently cause $Y$, without causing each other. Additionally, suppose the effects of both $A$ and $B$ on $Y$ are monotonic, meaning $\Pr(Y_{a_0}=y_1, Y_{a_1}=y_0) = \Pr(Y_{b_0}=y_1, Y_{b_1}=y_0) = 0$, and that there are no interactions, meaning $\Pr(Y_{a, b}=y_1) - \Pr(Y_{a', b}=y_1) = \Pr(Y_{a, b'}=y_1) - \Pr(Y_{a', b'}=y_1)$ and $\Pr(Y_{a, b}=y_1) - \Pr(Y_{a, b'}=y_1) = \Pr(Y_{a', b}=y_1) - \Pr(Y_{a', b'}=y_1)$. Under these conditions, without imposing any other assumptions, such as absence of confounding between $A$, $B$, and $Y$, the probability distribution $\Pr(Y_a)$, representing the outcome under different levels of treatment $A$, is not identifiable solely from the joint distribution $\Pr(Y_{a, b})$ and $\Pr(Y, A, B)$.  Finally, these quantities are not identifiable even if we relax the assumption of monotonicity.
\end{theorem}
\begin{proof}
    Under monotonicity and no interaction assumptions, there are only four possible functions from $A, B$ to $Y$: $Y = 1, Y = A, Y = B, Y = 0$. We construct two different models, $M_0$ and $M_1$ using those functions that induce the same observational and factorial distribution, but that diverge with respect to single-treatment quantities. 

    Consider models $M_0$ and $M_1$, where errors in both models are determined by two independent variables: a ternary $U_{aby}$ with probabilities $\Pr(U_{aby} = 0) = 0.1, \Pr(U_{aby} = 1) = 0.3, \Pr(U_{aby} = 2) = 0.6$ and a ternary uniform $U_y$ with $\Pr(U_y = u) = 1/3$ for any $u$. Let $A = 0$, if  $\Pr(U_{aby} = 0)$ and $A = 1$, if $\Pr(U_{aby} \neq 0)$. Let $B = 0$, if  $\Pr(U_{aby} \neq 2)$ and $B = 1$, if  $\Pr(U_{aby} = 2)$.

Let $Y_{M_0}$ be determined by the function:
    \begin{align*}
        Y_{M_0} = \begin{cases}
            0,\text{ if } & U_{aby} = 1, U_y = 0  \\
            1,\text{ if } & U_{aby} = 1, U_y = 1 \\
            B,\text{ if } & U_{aby} = 0, U_y = 0 \text{ or } U_{aby} = 2, U_y \neq 0 \\ 
            A,\text{ otherwise } &
        \end{cases}
    \end{align*}

    Let $Y_{M_1}$ be determined by the function:
    \begin{align*}
        Y_{M_1} = \begin{cases}
            0,\text{ if } & U_{aby} = 1, U_y = 0  \\
            1,\text{ if } & U_{aby} = 1, U_y = 1 \\
            B,\text{ if } & U_{aby} = 2, \\ 
            A,\text{ otherwise } &
        \end{cases}
    \end{align*}

Computation was omitted, but outputs are presented in Table~\ref{tab:probnointeractmono}, which show that both $\Pr_{M_0}(A,B,Y) = \Pr_{M_1}(A,B,Y)$ and $\Pr_{M_0}(Y_{a,b}) = \Pr_{M_1}(Y_{a,b})$ but $\Pr_{M_0}({Y_a}) \neq \Pr_{M_1}({Y_a})$. That suffices to show $\Pr({Y_a})$ is not identifiable.     
\end{proof}

\begin{table}[htbp]
\centering
\caption{Calculated probabilities for models $M_0$ and $M_1$ as presented in the counterexample detailed in Theorem~\ref{theorem:nointeractmono}. Despite both models being monotonic, they agree on all observational and factorial (2-treatment) quantities. However, they diverge regarding $\Pr(Y_{A=1})$ and $\Pr(Y_{A=0})$. That implies both quantities are not point-identifiable.}
\label{tab:probnointeractmono}
\begin{tabular}{@{}l|ccc@{}}
\toprule
\textbf{Probability} & \textbf{Model $M_0$} & \textbf{Model $M_1$} \\
\midrule
\multicolumn{3}{@{}l}{\textbf{Observational}} \\
\midrule
$\Pr(A = 0, B = 0, Y = 0)$ & 0.1 & 0.1 \\
$\Pr(A = 0, B = 1, Y = 0)$ & 0 & 0 \\
$\Pr(A = 1, B = 0, Y = 0)$ & 0 & 0 \\
$\Pr(A = 1, B = 1, Y = 0)$ & 0 & 0 \\
$\Pr(A = 0, B = 0, Y = 1)$ & 0.1 & 0.1 \\
$\Pr(A = 0, B = 1, Y = 1)$ & 0.2 & 0.2 \\
$\Pr(A = 1, B = 0, Y = 1)$ & 0 & 0 \\
$\Pr(A = 1, B = 1, Y = 1)$ & 0.6 & 0.6 \\
\midrule
\multicolumn{3}{@{}l}{\textbf{2-Treatment}} \\
\midrule
$\Pr(Y_{A=0, B = 0} = 1)$ & 0.1 & 0.1 \\
$\Pr(Y_{A=0, B = 1} = 1)$ & 0.4 & 0.4 \\
$\Pr(Y_{A=1, B = 0} = 1)$ & 0.6 & 0.6 \\
$\Pr(Y_{A=1, B = 1} = 1)$ & 0.9 & 0.9 \\
\midrule
\multicolumn{3}{@{}l}{\textbf{1-Treatment}} \\
\midrule
$\Pr(Y_{A=1}=1)$ & 0.8 & 0.9 \\
$\Pr(Y_{A=0}=1)$ & 0.3 & 0.4 \\
\bottomrule
\end{tabular}
\end{table}

 In the scenario outlined in Theorem~\ref{theorem:nointeractmono}, if we remove the assumption of monotonicity, $\Pr(Y_a)$ remains unidentifiable solely from the joint distribution $\Pr(Y_{a, b})$ and $\Pr(Y, A, B)$. This is due to the fact that the non-identification result remains in less restrictive cases.

\begin{theorem}
    \label{theorem:nointeractonlyfact}
  Consider treatments $A$ and $B$, along with an outcome $Y$. Assume that both $A$ and $B$ independently cause $Y$, without causing each other. Additionally, suppose the effects of both $A$ and $B$ on $Y$ are monotonic, meaning $\Pr(Y_{a_0}=y_1, Y_{a_1}=y_0) = \Pr(Y_{b_0}=y_1, Y_{b_1}=y_0) = 0$, and that there are no interactions, meaning $\Pr(Y_{a, b}=y_1) - \Pr(Y_{a', b}=y_1) = \Pr(Y_{a, b'}=y_1) - \Pr(Y_{a', b'}=y_1)$ and $\Pr(Y_{a, b}=y_1) - \Pr(Y_{a, b'}=y_1) = \Pr(Y_{a', b}=y_1) - \Pr(Y_{a', b'}=y_1)$. Under these conditions, without imposing any other assumptions, such as absence of confounding between $A$, $B$, and $Y$, the probability distribution $\Pr(Y_a)$, representing the outcome under different levels of treatment $A$, is not identifiable solely from the factorial distribution $\Pr(Y_{a, b})$ or from observational distribution $\Pr(y,a,b)$ alone. Furthermore, only in the last case, (observational distribution alone), the ATE remains unidentifiable. 
\end{theorem}
\begin{proof}
    Consider two models $M_0$ and $M_1$ on binary $A, B$ and $Y$. Assume we fix the probability under interventions (factorial data) on $B$ and $A$, $\Pr_{M=0}(Y_{a,b}) =  \Pr_{M=1}(Y_{a,b})$ and that monotonicity is satisfied.    As $B$ is independent of $Y_{a,b}$, set $\Pr_{M=0}(b) \neq \Pr_{M=0}(b)$.    

Now assume we rather have only observational data, $\Pr(y,a,b)$. In this case, as we have unobserved confounding, $Y_{a}$ or the ATE would not be identifiable. Those results remain, even if we assume monotonicity, provided that non-identification results persist in less restrictive scenarios.
    
\end{proof}

\section{Parametric Identification of Single-Treatment Estimands Using Factorial Data}
\label{proof:parametric}

\begin{theorem}[AS is sufficient for identifiability of the ATE]
\label{theorem:as}
Suppose $y_i = f(a_i, u_i) + g(b_i, u_i)$ with unobserved $u_i$ for arbitrary fixed functions $f, g$ is a model of the real-world and assume statistical dependence between $a_i$ and $b_i$ and $u_i$. 

Suppose one has data on $\Pr(Y(a,b))$, for all states $a,b,y$. Then the ATE of $A$ on $Y$ is identifiable by 
\begin{equation}
\label{equation:amce}
    \sum_b (\EX(Y_{a,b}) - \EX(Y_{a',b})) f(b), \text{ with uniform } f(b)
\end{equation}

\end{theorem}
\begin{proof}
Suppose $y_i = f(a_i, u_i) + g(b_i, u_i)$ for arbitrary functions $f, g$, then $E(Y_{a}) = \EX(f(a, u_i)) +\EX(g(b_i, u_i))$ and $\EX(Y_{a'}) = \EX(f(a', u_i)) + \EX(g(b_i))$. Then the ATE is $\EX(Y_{a} - Y_{a'}) = \EX(f(a, u_i) - f(a', u_i))$.

So it is sufficient to identify $\EX(f(a, u_i) - f(a', u_i))$. Using the factorial data we have the following identifiable equality for any $b \in \supp(B)$:
\begin{align*}
    \EX(Y_{a,b}) = \EX(f(a, u_i)) + \EX(g(b, u_i)),\\
\end{align*}

Then $\EX(Y_{a} - Y_{a'}) = \EX(f(a, u_i) - f(a', u_i))$ is identifiable by either $\EX(Y_{a, b}) - \EX(Y_{a', b})$ for any $b$. That implies that the 
for a uniform distribution $f(b)$, the ATE is also identified by $\sum_{b \in \supp(B)} (\EX(Y_{a_1,b}) - \EX(Y_{a_0,b})) f(b)$.

\end{proof}

The separability criterion described above does not accept interactions. However, for linear models with interactions, identifiability is still possible. 

\begin{theorem}[Sufficiency of Interactive Linearity For Single-treatment Quantities]
\label{theorem:lininteract}
    Suppose $y_i = \beta_0 + \beta_1 a_i + \beta_2 b_i + \beta_3 a_i b_i + u_i$ with unobserved $u_i$ is a model of the real-world, discrete $b_i$ and $a_i$, and assume statistical dependence between $a_i$ and $b_i$ and $u_i$. Then the expected outcome of an intervention $A = a$ is identifiable using the observational quantity $\Pr(B = b)$ and the factorial distribution $\EX(Y_{a,b})$ by:
\begin{equation}
\label{equation:interactivesingle}
    \EX(Y_a) = \sum_{b} \EX(Y_{a, b}) \Pr(b)
\end{equation}
    and the ATE is by the quantity:
    \begin{align*}
    \EX(Y_a) - \EX(Y_{a'}) = \sum_{b} (\EX(Y_{a, b}) - \EX(Y_{a', b}))\Pr(b)
\end{align*}
\end{theorem}
\begin{proof}
We make the simplifying assumption that $\EX(u_i) = 0$. Then we can rewrite $\EX(Y_a)$ as:
\begin{align*}
    \EX(Y_a)  = \\
    \beta_0 + \beta_1 \cdot a +  \beta_2 \EX(b_i) + \beta_3 \EX(b_i) \cdot a \\
\end{align*}

As $\Pr(b)$ is directly identifiable from the observational data and $\EX(Y_{a,b})$ from the factorial data, then:
\begin{align*}
    \sum_{b} \EX(Y_{a, b}) \Pr(b) = \\
    \sum_{b} \Pr(b) (\beta_0 + \beta_1 a + \beta_2 b + \beta_3 a b ) = \\
     \beta_0 \sum_{b} \Pr(b) +  \beta_1 a \sum_{b} \Pr(b)  + \beta_2 \sum_{b} b  \Pr(b) + \beta_3 a \sum_{b} b \Pr(b) =  \\
     \beta_0 + \beta_1 a + \beta_2 \EX(b_i) + a \beta_3 \EX(b_i) = \\
     \EX(Y_a)
\end{align*}

That confirms that the identification of $\EX(Y_a) = \sum_b \EX(Y_{a, b}) \Pr(b)$ and implies 
that the ATE is identifiable by 
\begin{align*}
    \EX(Y_{a} - Y_{a'}) = \sum_b ( \EX(Y_{a, b}) -  \EX(Y_{a', b})) \Pr(b)
\end{align*}
\end{proof}

\begin{corollary}[Linearity is Sufficient for Identifiability]
    \label{corollary:linearity}
Suppose $y_i = \beta_0 + \beta_1 a_i + \beta_2 b_i + u_i$ with unobserved $u_i$ is a model of the real-world and assume statistical dependence between $a_i$ and $b_i$ and $u_i$. Then the ATE of $A$ on $Y$ is identifiable using factorial data by the quantity in equation~\ref{equation:amce} as that is a case of AS (Theorem~\ref{theorem:as}). Additionally, linearity is also a case of interactive linearity (Theorem~\ref{theorem:lininteract}), so single-treatment quantities are identifiable by Equation~\ref{equation:interactivesingle}. Finally, observational data alone is insufficient for identifiability given the standard omitted variable bias (unobserved confounding). 
\end{corollary}

\section{Partial identification results}

\begin{theorem}[Bounds only with factorial distribution]
\label{theorem:boundsonlyfact}
Suppose $A, B, Y$ are binary random variables and we have only factorial data -- $\Pr(Y_{a,b})$. Then the sharp bounds for the expected value of $Y$ given intervention on $A$, $\Pr(Y_a)$ are:
\begin{align*}
\max\{0, 1 - \sum_b \Pr(Y_{ab} \neq y)\} \leq \Pr(Y_a) \leq \min\{\sum_b \Pr(Y_{ab} = y), 1\}
\end{align*}

and the bounds for the ATE ( $\EX(Y_{a_1} - Y_{a_0}$) are:
\begin{align*}
\max \begin{cases}
-1 \\
 - \EX(Y_{a_0b_0}  + Y_{a_0b_1}) \\
 \EX(Y_{a_1b_0}  + Y_{a_1 b_1}) - 2 \\
  \EX(Y_{a_1b_0} + Y_{a_1b_1} - Y_{a_0b_0} - Y_{a_0b_1} ) - 1
    \end{cases}  \leq 
    \text{ATE} \leq 
    \min 
    \begin{cases}
    1 \\
 2 -    \EX(Y_{a_0b_1}  + Y_{a_0b_0}) \\
   \EX(Y_{a_1b_0} + Y_{a_1b_1} )  \\ 
   1 + \EX(Y_{a_1b_0} + Y_{a_1b_1} - Y_{a_0b_0} - Y_{a_0b_1})
    \end{cases}
\end{align*}

\end{theorem}

\begin{proof}

This result follows from the solution of the linear program detailed below. Because that solution is exact, these are the sharp bounds for the problem.
\end{proof}

\begin{theorem}[Bounds with both observational and factorial distributions]
\label{theorem:boundsmain}
Suppose $A, B, Y$ are binary random variables and we have both factorial -- $\Pr(Y_{a,b})$ and observational data -- $\Pr(A,B,Y)$. Let $x' = 1 - x$ for any $x$. Then the sharp bounds for a quantity such as the expected value of $Y$ given intervention on $A$, $E(Y_a)$ are:
\begin{align*} \max  
\begin{cases} 0, \\
\Pr(Y_{ab'} = y) - \Pr(a',b) - \Pr(a,b,y') , \\
\Pr(a,y) , \\
\Pr(Y_{ab} = y') - \Pr(b',y') -  \Pr(a',b',y), \\
\Pr(Y_{ab'} = y) - \Pr(Y_{ab} = y')
\end{cases} \leq \EX(Y_a)  \leq \min
\begin{cases} 1, \\ 
\Pr(a',b,y') + \Pr(b,y) + \Pr(Y_{ab'} = y), \\
1 - \Pr(a,y'), \\
\Pr(Y_{ab'} = y) + \Pr(Y_{ab} = y), \\
\Pr(a',b') + \Pr(a,b',y) + \Pr(Y_{ab} = y)
\end{cases}
\end{align*}

and the sharp lower and upper bounds for the ATE are respectively:
\begin{align*}
\max \begin{cases}
- \Pr(a_1,y_0) - \Pr(a_0, y_1)  \\
\Pr(a_1,y_1) - \Pr(Y_{a_0b_0} = y_1) - \Pr(Y_{a_0b_1} = y_1)  \\
\Pr(a_1,b_1) - \Pr(a_1,y_0) - \Pr(a_0,b_0, y_1) - \Pr(Y_{a_0b_1} = y_1) \\
\Pr(a_1,b_0) - \Pr(a_1,y_0) - \Pr(a_0,b_1, y_1) - \Pr(Y_{a_0b_0} = y_1) \\
\Pr(a_0, y_0) - \Pr(Y_{a_1b_0} = y_0) - \Pr(Y_{a_1b_1} = y_0) \\
 \Pr(Y_{a_1b_0} = y_1) - \Pr(Y_{a_1b_1} = y_0) - \Pr(Y_{a_0b_0} = y_1) - \Pr(Y_{a_0b_1} = y_1) \\
  \Pr(Y_{a_1b_0} = y_1) - \Pr(Y_{a_1b_1} = y_0) - \Pr(Y_{a_0b_1} = y_1) - \Pr(a_0, b_0, y_1) - \Pr(a_1, b_0) \\
    \Pr(Y_{a_1b_0} = y_1) - \Pr(Y_{a_1b_1} = y_0) - \Pr(Y_{a_0b_0} = y_1) - \Pr(a_0, b_1, y_1) - \Pr(a_1, b_1) \\
    \Pr(a_0, b_1) - \Pr(a_1, b_0, y_0) - \Pr(a_0, y_1) - \Pr(Y_{a_1b_1} = y_0) \\ 
    \Pr(Y_{a_1b_1} = y_1) - \Pr(Y_{a_0b_0} = y_1) - \Pr(Y_{a_0b_1} = y_1) - \Pr(a_1, b_0, y_0) - \Pr(a_0, b_0) \\
\Pr(Y_{a_1b_1} = y_1) - \Pr(Y_{a_0b_1} = y_1)  - \Pr(b_0) - \Pr(a_1, b_0, y_0) - \Pr(a_0, b_0, y_1) \\
\Pr(a_1, b_0, y_1) + \Pr(a_0, b_1, y_0) - \Pr(Y_{a_1b_1} = y_0) - \Pr(Y_{a_0b_0} = y_1) \\
\Pr(a_0, b_0) - \Pr(a_1, b_1, y_0) - \Pr(a_0, y_1) -  \Pr(Y_{a_1b_0} = y_0) \\
\Pr(Y_{a_1b_0}=y_1) - \Pr(Y_{a_0b_0}=y_1) - \Pr(Y_{a_0b_1}=y_1) - \Pr(a_0,b_1) - \Pr(a_1,b_1,y_0) \\
\Pr(a_1,b_1,y_1) + \Pr(a_0,b_0,y_0) - \Pr(Y_{a_1b_0}=y_0) - \Pr(Y_{a_0b_1}=y_1) \\
\Pr(Y_{a_1b_0}=y_1) - \Pr(Y_{a_0b_1}=y_1) - \Pr(a_0,b_1) - \Pr(a_1,b_1,y_0) - \Pr(a_0,b_1,y_1) - \Pr(a_1,b_1) 
    \end{cases} 
    \end{align*}
    and \begin{align*} 
    \min  \begin{cases}
        \Pr(a_1,y_1) + \Pr(a_0, y_0) \\ 
        \Pr(Y_{a_0b_0}=y_0) + \Pr(Y_{a_0b_1}=y_0) - \Pr(a_1,y_0) \\
        \Pr(Y_{a_0b_1}=y_0)  + \Pr(a_0,b_0,y_0) + \Pr(a_1,b_0) - \Pr(a_1,y_0)\\
        \Pr(Y_{a_0b_0}=y_0) + \Pr(a_0,b_1,y_0) + \Pr(a_1,b_1) - \Pr(a_1,y_0) \\
        \Pr(Y_{a_1b_0}=y_1) + \Pr(Y_{a_1b_1}=y_1) - \Pr(a_0,y_1) \\
        \Pr(Y_{a_1b_0}=y_1) + \Pr(Y_{a_1b_1}=y_1) + \Pr(Y_{a_0b_0}=y_0) - \Pr(Y_{a_0b_1}=y_1)
\\
        \Pr(Y_{a_1b_0}=y_1) + \Pr(Y_{a_1b_1}=y_1) - \Pr(Y_{a_0b_1}=y_1) + \Pr(b_0,y_0) + \Pr(a_1,b_0,y_1)    \\
        \Pr(Y_{a_1b_0}=y_1) + \Pr(Y_{a_1b_1}=y_1) - \Pr(Y_{a_0b_0}=y_1) + \Pr(b_1,y_0) + \Pr(a_1,b_1,y_1)    \\
        \Pr(Y_{a_1b_1}=y_1) + \Pr(a_1,b_0,y_1) + \Pr(a_0,b_0,y_0) - \Pr(a_0,b_1, y_1) \\
        \Pr(Y_{a_1b_1}=y_1) + \Pr(Y_{a_0b_1}=y_0) - \Pr(Y_{a_0b_0}=y_1) + \Pr(a_0,b_0) + \Pr(a_1,b_0,y_1) \\
        \Pr(Y_{a_1b_1}=y_1) - \Pr(Y_{a_0b_1}=y_1) + \Pr(a_0,b_0) +  \Pr(b_0,y_0) + 2 \Pr(a_1,b_0,y_1)  \\
        \Pr(Y_{a_1b_1}=y_1) + \Pr(Y_{a_0b_0}=y_0) -  \Pr(a_1,b_0,y_0) - \Pr(a_0,b_1,y_1)
    \\
        \Pr(Y_{a_1b_0}=y_1) -  \Pr(a_0,b_0) + \Pr(a_1,b_1,y_1) + \Pr(a_0,y_0) \\ 
        \Pr(Y_{a_1b_0}=y_1) - \Pr(Y_{a_0b_0}=y_1) + \Pr(Y_{a_0b_1}=y_0) + \Pr(a_0,b_1) +   \Pr(a_1,b_1,y_1) \\ 
        \Pr(Y_{a_1b_0}=y_1) - \Pr(Y_{a_0b_1}=y_1) +  \Pr(a_0,b_1) + \Pr(a_1,y_1) + \Pr(b_0,y_0) \\ 
        \Pr(Y_{a_1b_0}=y_1) - \Pr(Y_{a_0b_0}=y_1) +  \Pr(a_0,b_1) +  \Pr(b_1,y_0) + 2 \Pr(a_1,b_1, y_1)
    \end{cases}
    \end{align*}
\end{theorem}

\begin{proof}

This result follows from the solution of the linear program detailed below. Because that solution is exact, these are the sharp bounds for the problem.
\end{proof}

\begin{theorem}[Bounds with monotonicity and using only factorial distribution]
\label{thm:boundsonlyfactmono}
Suppose $A, B, Y$ are binary random variables and we have only factorial data -- $\Pr(Y_{a,b})$. Assume the causal effects of $A$ and $B$ on $Y$ are monotonic. Then the sharp bounds both for the expected value of $Y$ given intervention on $A$, $\Pr(Y_a = y_1)$:
\begin{align*}
0 \leq \Pr(Y_{a_1} = y_1) \leq \min\{\sum_b \Pr(y_{ab}), 1\} \\
\end{align*}
, and for the ATE are:
\begin{align*}   
\max \begin{cases}
    0, \\
    - \Pr(Y_{a0b0} = y_0)  - \Pr(Y_{a0b1} = y_1)  + \Pr(Y_{a1b0} = y_0)  - \Pr(Y_{a1b1} = y_0)  \\
    - \Pr(Y_{a0b0} = y_1)  - \Pr(Y_{a0b1} = y_0)  - \Pr(Y_{a1b0} = y_0)  + \Pr(Y_{a1b1} = y_0)  \\
    - \Pr(Y_{a0b0} = y_`)  - \Pr(Y_{a0b1} = y_1)  + \Pr(Y_{a1b0} = y_1)  - \Pr(Y_{a1b1} = y_0)  \\
\end{cases} \leq \text{ ATE }\end{align*} 
and
\begin{align*}   
\text{ ATE } \leq \min \begin{cases}
1 \\
 \Pr(Y_{a0b0} = y_0) -  \Pr(Y_{a0b1} = y_1)  + \Pr(Y_{a1b0} = y_1)  + \Pr(Y_{a1b1} = y_1)  \\
   \Pr(Y_{a0b0} = y_0)  + \Pr(Y_{a1b1} = y_1)   \\
      \Pr(Y_{a0b1} = y_0)  + \Pr(Y_{a1b0} = y_1)   \\
      \Pr(Y_{a1b0} = y_1)  + \Pr(Y_{a1b1} = y_1)   \\
   \Pr(Y_{a0b0} = y_0)  + \Pr(Y_{a0b1} = y_)   \\
\end{cases}
\end{align*}
\end{theorem}

\begin{proof}
This result follows from the solution of the linear program detailed below. Because that solution is exact, these are the sharp bounds for the problem.
\end{proof}

\begin{theorem}[Bounds with monotonicity and using both distributions]
Suppose $A, B, Y$ are binary random variables and we have both factorial -- $\Pr(Y_{a,b})$ and observational data -- $\Pr(A,B,Y)$. Assume the causal effects of $A$ and $B$ on $Y$ are monotonic. Then the sharp bounds for a quantity such as the expected value of $Y$ given intervention on $A$, $E(Y_{a_1})$ are:
\begin{align*} \max  
\begin{cases} 
0\\
- \Pr(Y_{a_0b_0} = y_1) -  \Pr(Y_{a_0b_1} = y_1) +  \Pr(Y_{a_1b_0} = y_1) - \Pr(Y_{a_1b_1} = y_0)\\
- \Pr(Y_{a_0b_0} = y_0) - \Pr(Y_{a_0b_1} = y_1) + \Pr(Y_{a_1b_0} = y_0) - \Pr(Y_{a_1b_1} = y_0)\\
 \Pr(Y_{a_0b_0} = y_0) - \Pr(Y_{a_0b_1} = y_0) -  \Pr(Y_{a_1b_0} = y_0) - \Pr(Y_{a_1b_1} = y_1)\\
\end{cases} \leq \EX(Y_{a_1})
\end{align*} and
\begin{align*} \EX(Y_{a_1}) \leq \min
\begin{cases} 
1 \\
 \Pr(Y_{a_0b_0} = y_0) +  \Pr(Y_{a_1b_1} = y_1)\\
- \Pr(Y_{a_0b_0} = y_1)  + \Pr(Y_{a_0b_1} = y_0)  + \Pr(Y_{a_1b_0} = y_1) + \Pr(Y_{a_1b_1} = y_1)\\
 \Pr(Y_{a_1b_0} = y_1) + \Pr(Y_{a_1b_1} = y_1)\\
 \Pr(Y_{a_0b_1} = y_0)  + \Pr(Y_{a_1b_0} = y_1)\\
\Pr(Y_{a_0b_0} = y_0) - \Pr(Y_{a_0b_1} = y_1) + \Pr(Y_{a_1b_0} = y_0) + \Pr(Y_{a_1b_0} = y_1)\\
\end{cases}
\end{align*}
and the sharp lower and upper bounds for the ATE are respectively:
\begin{align*}
\max \begin{cases}
0\\
\Pr(b_0, y_1) - \Pr(Y_{a_0b_0} = y_1)\\
- \Pr(y_1) - \Pr(a_0, b_1, y_0)   + \Pr(Y_{a_1b_0} = y_1)\\
- \Pr(a_0, b_1) - \Pr(a_1, b_1, y_1) - \Pr(Y_{a_0b_0} = y_1) + \Pr(Y_{a_1b_0} = y_1)\\
\Pr(a_0, b_0, y_1) + \Pr(b_1, y_1) - \Pr(Y_{a_0b_1} = y_1)\\
\Pr(y_1) + \Pr(a_0, b_0, y_1)   - \Pr(Y_{a_0b_0} = y_1) - \Pr(Y_{a_0b_1} = y_1)\\
\Pr(a_0, b_0, y_1) - \Pr(a_0, b_1, y_0) - \Pr(Y_{a_0b_0} = y_1) - \Pr(Y_{a_0b_1} = y_1) + \Pr(Y_{a_1b_0} = y_1)\\
- \Pr(a_0, b_1, y_0) - \Pr(a_1, b_0, y_1) - \Pr(Y_{a_0b_1} = y_1) + \Pr(Y_{a_1b_0} = y_1)\\
- \Pr(b_0) - \Pr(b_1, y_1)  + \Pr(Y_{a_1b_1} = y_1)\\
 - \Pr(b_1, y_1) - \Pr(b_0, y_0)  - \Pr(Y_{a_0b_0} = y_1) + \Pr(Y_{a_1b_1} = y_1)\\
- \Pr(y_1) + \Pr(a_1, b_1, y_0) -  \Pr(Y_{a_1b_0} = y_0) + \Pr(Y_{a_1b_1} = y_1)\\
 -  \Pr(b_1, y_1) +  \Pr(a_1, b_1, y_0) - \Pr(Y_{a_0b_0} = y_1) - \Pr(Y_{a_1b_0} = y_0) + \Pr(Y_{a_1b_1} = y_1)\\
- \Pr(b_0, y_0) - \Pr(a_1, b_0, y_1) - \Pr(Y_{a_0b_1} = y_1) + \Pr(Y_{a_1b_1} = y_1)\\
- \Pr(a_0, b_0) - \Pr(a_1, b_0, y_0) - \Pr(Y_{a_0b_0} = y_1) - \Pr(Y_{a_0b_1} = y_1) + \Pr(Y_{a_1b_1} = y_1)\\
 \Pr(a_1, b_0, y_1) + \Pr(a_1, b_0, y_1)  - \Pr(Y_{a_0b_0} = y_1) - \Pr(Y_{a_0b_1} = y_1) - \Pr(Y_{a_1b_0} = y_0) + \Pr(Y_{a_1b_1} = y_1)\\
- \Pr(a_0) - \Pr(a_1, b_0) - \Pr(a_1, y_1)  - \Pr(Y_{a_0b_1} = y_1) + \Pr(Y_{a_1b_0} = y_1) + \Pr(Y_{a_1b_1} = y_1)
\end{cases}
\end{align*}
and
\begin{align*}
\min \begin{cases}
 - \Pr(a_0, b_0, y_1) + \Pr(a_0, b_1, y_0) + \Pr(a_1, b_1, y_1) + \Pr(Y_{a_1b_0} = y_1)\\
 \Pr(a_0, b_0, y_0) + \Pr(a_1, b_0, y_1) + \Pr(a_1, b_1, y_0) + \Pr(Y_{a_0b_1} = y_0)\\
 \Pr(a_0, b_0, y_0) + \Pr(b_0) - \Pr(Y_{a_0b_1} = y_1) + \Pr(Y_{a_1b_1} = y_1)\\
 \Pr(a_0, y_0) + \Pr(a_1,  y_1) \\
 \Pr(a_0, y_0) +  \Pr(b_1, y_1) + \Pr(a_1, b_0)  - \Pr(Y_{a_0b_1} = y_1) + \Pr(Y_{a_1b_0} = y_1)\\
 \Pr(b_1) + \Pr(a_1, b_1, y_1) - \Pr(Y_{a_0b_0} = y_1) + \Pr(Y_{a_1b_0} = y_1)\\
 - \Pr(a_0, b_1, y_0) - \Pr(a_1, y_0)  + \Pr(Y_{a_0b_0} = y_0) + \Pr(Y_{a_0b_1} = y_0)\\
 \Pr(b_0, y_0)  - \Pr(Y_{a_0b_1} = y_1) + \Pr(Y_{a_1b_0} = y_1) + \Pr(Y_{a_1b_1} = y_1)\\
 \Pr(a_0, b_0, y_0) - \Pr(a_0, b_1, y_1) + \Pr(Y_{a_1b_1} = y_1)\\
 - \Pr(a_1, b_0)  + \Pr(a_1, b_1)  + \Pr(Y_{a_0b_0} = y_0)\\
 \Pr(a_0, b_0) -  \Pr(a_0, b_1, y_0)  + \Pr(Y_{a_0b_0} = y_0) - \Pr(Y_{a_0b_1} = y_1) + \Pr(Y_{a_1b_1} = y_1)\\
 \Pr(a_0, b_0) + \Pr(a_1, b_1) - \Pr(Y_{a_0b_0} = y_1) + \Pr(Y_{a_1b_1} = y_1)\\
 \Pr(a_1, b_1, y_1) - \Pr(Y_{a_0b_0} = y_1) + \Pr(Y_{a_0b_1} = y_0) + \Pr(Y_{a_1b_0} = y_1)\\
 - \Pr(a_0, y_1) - \Pr(a_1, b_0, y_1) + \Pr(Y_{a_1b_0} = y_1) + \Pr(Y_{a_1b_1} = y_1)\\
 - \Pr(a_1, b_0, y_1) + \Pr(a_1, b_1) - \Pr(Y_{a_0b_0} = y_1) + \Pr(Y_{a_1b_0} = y_1) + \Pr(Y_{a_1b_1} = y_1)\\
\Pr(a_0, b_0) + \Pr(b_1, y_1) + \Pr(a_1, y_0) - \Pr(Y_{a_0b_0} = y_1) - \Pr(Y_{a_0b_1} = y_1) + \Pr(Y_{a_1b_0} = y_1) + \Pr(Y_{a_1b_1} = y_1)\\
\end{cases}
\end{align*}

\end{theorem}
\begin{proof}
This result follows from the solution of the linear program detailed below. Because that solution is exact, these are the sharp bounds for the problem.
\end{proof}

\section{Linear programs}
\label{linearprogram}
This is the linear program for deriving sharp bounds for both estimands $\Pr(Y_{a_1} = y_1)$  and the ATE using data on $\Pr(Y,A,B)$  and $\Pr(Y_{a,b})$.
Each probabilistic quantity is transformed into principal strata, which will become variables of the program \citep{duarte2023automated}. Data and axioms of probability become constraints, and the estimand, the objective function. The program might be solved using a symbolic or numeric linear program solver.

\begin{itemize}
    \item Variables: $[a_0b_0y_{0000}, ..., a_1b_1y_{1111}]$
    \item Constraints:
    \begin{itemize}
        \item Observational distribution: 
        \begin{itemize}
            \item $a_{0}b_{0}y_{0000} + a_{0}b_{0}y_{0001} + a_{0}b_{0}y_{0010} + a_{0}b_{0}y_{0011} + a_{0}b_{0}y_{0100} + a_{0}b_{0}y_{0101} + a_{0}b_{0}y_{0110} + a_{0}b_{0}y_{0111} = \Pr(a_0, b_0, y_0)$
\item $a_{0}b_{0}y_{1000} + a_{0}b_{0}y_{1001} + a_{0}b_{0}y_{1010} + a_{0}b_{0}y_{1011} + a_{0}b_{0}y_{1100} + a_{0}b_{0}y_{1101} + a_{0}b_{0}y_{1110} + a_{0}b_{0}y_{1111} = \Pr(a_0, b_0, y_1) $
\item $a_{0}b_{1}y_{0000} + a_{0}b_{1}y_{0001} + a_{0}b_{1}y_{0010} + a_{0}b_{1}y_{0011} + a_{0}b_{1}y_{1000} + a_{0}b_{1}y_{1001} + a_{0}b_{1}y_{1010} + a_{0}b_{1}y_{1011} = \Pr(a_0, b_1, y_0) $
\item $a_{0}b_{1}y_{0100} + a_{0}b_{1}y_{0101} + a_{0}b_{1}y_{0110} + a_{0}b_{1}y_{0111} + a_{0}b_{1}y_{1100} + a_{0}b_{1}y_{1101} + a_{0}b_{1}y_{1110} + a_{0}b_{1}y_{1111} = \Pr(a_0, b_1, y_1)$
\item $a_{1}b_{0}y_{0000} + a_{1}b_{0}y_{0001} + a_{1}b_{0}y_{0100} + a_{1}b_{0}y_{0101} + a_{1}b_{0}y_{1000} + a_{1}b_{0}y_{1001} + a_{1}b_{0}y_{1100} + a_{1}b_{0}y_{1101} = \Pr(a_1, b_0, y_0) $
\item $a_{1}b_{0}y_{0010} + a_{1}b_{0}y_{0011} + a_{1}b_{0}y_{0110} + a_{1}b_{0}y_{0111} + a_{1}b_{0}y_{1010} + a_{1}b_{0}y_{1011} + a_{1}b_{0}y_{1110} + a_{1}b_{0}y_{1111} = \Pr(a_1, b_0, y_1) $
\item $a_{1}b_{1}y_{0000} + a_{1}b_{1}y_{0010} + a_{1}b_{1}y_{0100} + a_{1}b_{1}y_{0110} + a_{1}b_{1}y_{1000} + a_{1}b_{1}y_{1010} + a_{1}b_{1}y_{1100} + a_{1}b_{1}y_{1110} = \Pr(a_1, b_1, y_0) $ 
\item $a_{1}b_{1}y_{0001} + a_{1}b_{1}y_{0011} + a_{1}b_{1}y_{0101} + a_{1}b_{1}y_{0111} + a_{1}b_{1}y_{1001} + a_{1}b_{1}y_{1011} + a_{1}b_{1}y_{1101} + a_{1}b_{1}y_{1111} = \Pr(a_1, b_1, y_1) $
        \end{itemize}
        \item Factorial distribution:
        \begin{itemize}
        \item $a_{0}b_{0}y_{0000} + a_{0}b_{0}y_{0001} + a_{0}b_{0}y_{0010} + a_{0}b_{0}y_{0011} + a_{0}b_{0}y_{0100} + a_{0}b_{0}y_{0101} + a_{0}b_{0}y_{0110} + a_{0}b_{0}y_{0111} + a_{0}b_{1}y_{0000} + a_{0}b_{1}y_{0001} + a_{0}b_{1}y_{0010} + a_{0}b_{1}y_{0011} + a_{0}b_{1}y_{0100} + a_{0}b_{1}y_{0101} + a_{0}b_{1}y_{0110} + a_{0}b_{1}y_{0111} + a_{1}b_{0}y_{0000} + a_{1}b_{0}y_{0001} + a_{1}b_{0}y_{0010} + a_{1}b_{0}y_{0011} + a_{1}b_{0}y_{0100} + a_{1}b_{0}y_{0101} + a_{1}b_{0}y_{0110} + a_{1}b_{0}y_{0111} + a_{1}b_{1}y_{0000} + a_{1}b_{1}y_{0001} + a_{1}b_{1}y_{0010} + a_{1}b_{1}y_{0011} + a_{1}b_{1}y_{0100} + a_{1}b_{1}y_{0101} + a_{1}b_{1}y_{0110} + a_{1}b_{1}y_{0111} = \Pr(Y_{a_0b_0}=y_0) $
\item $a_{0}b_{0}y_{1000} + a_{0}b_{0}y_{1001} + a_{0}b_{0}y_{1010} + a_{0}b_{0}y_{1011} + a_{0}b_{0}y_{1100} + a_{0}b_{0}y_{1101} + a_{0}b_{0}y_{1110} + a_{0}b_{0}y_{1111} + a_{0}b_{1}y_{1000} + a_{0}b_{1}y_{1001} + a_{0}b_{1}y_{1010} + a_{0}b_{1}y_{1011} + a_{0}b_{1}y_{1100} + a_{0}b_{1}y_{1101} + a_{0}b_{1}y_{1110} + a_{0}b_{1}y_{1111} + a_{1}b_{0}y_{1000} + a_{1}b_{0}y_{1001} + a_{1}b_{0}y_{1010} + a_{1}b_{0}y_{1011} + a_{1}b_{0}y_{1100} + a_{1}b_{0}y_{1101} + a_{1}b_{0}y_{1110} + a_{1}b_{0}y_{1111} + a_{1}b_{1}y_{1000} + a_{1}b_{1}y_{1001} + a_{1}b_{1}y_{1010} + a_{1}b_{1}y_{1011} + a_{1}b_{1}y_{1100} + a_{1}b_{1}y_{1101} + a_{1}b_{1}y_{1110} + a_{1}b_{1}y_{1111} = \Pr(Y_{a_0b_0}=y_1) $
\item $a_{0}b_{0}y_{0000} + a_{0}b_{0}y_{0001} + a_{0}b_{0}y_{0010} + a_{0}b_{0}y_{0011} + a_{0}b_{0}y_{1000} + a_{0}b_{0}y_{1001} + a_{0}b_{0}y_{1010} + a_{0}b_{0}y_{1011} + a_{0}b_{1}y_{0000} + a_{0}b_{1}y_{0001} + a_{0}b_{1}y_{0010} + a_{0}b_{1}y_{0011} + a_{0}b_{1}y_{1000} + a_{0}b_{1}y_{1001} + a_{0}b_{1}y_{1010} + a_{0}b_{1}y_{1011} + a_{1}b_{0}y_{0000} + a_{1}b_{0}y_{0001} + a_{1}b_{0}y_{0010} + a_{1}b_{0}y_{0011} + a_{1}b_{0}y_{1000} + a_{1}b_{0}y_{1001} + a_{1}b_{0}y_{1010} + a_{1}b_{0}y_{1011} + a_{1}b_{1}y_{0000} + a_{1}b_{1}y_{0001} + a_{1}b_{1}y_{0010} + a_{1}b_{1}y_{0011} + a_{1}b_{1}y_{1000} + a_{1}b_{1}y_{1001} + a_{1}b_{1}y_{1010} + a_{1}b_{1}y_{1011} =  \Pr(Y_{a_0b_1}=y_0) $
\item $a_{0}b_{0}y_{0100} + a_{0}b_{0}y_{0101} + a_{0}b_{0}y_{0110} + a_{0}b_{0}y_{0111} + a_{0}b_{0}y_{1100} + a_{0}b_{0}y_{1101} + a_{0}b_{0}y_{1110} + a_{0}b_{0}y_{1111} + a_{0}b_{1}y_{0100} + a_{0}b_{1}y_{0101} + a_{0}b_{1}y_{0110} + a_{0}b_{1}y_{0111} + a_{0}b_{1}y_{1100} + a_{0}b_{1}y_{1101} + a_{0}b_{1}y_{1110} + a_{0}b_{1}y_{1111} + a_{1}b_{0}y_{0100} + a_{1}b_{0}y_{0101} + a_{1}b_{0}y_{0110} + a_{1}b_{0}y_{0111} + a_{1}b_{0}y_{1100} + a_{1}b_{0}y_{1101} + a_{1}b_{0}y_{1110} + a_{1}b_{0}y_{1111} + a_{1}b_{1}y_{0100} + a_{1}b_{1}y_{0101} + a_{1}b_{1}y_{0110} + a_{1}b_{1}y_{0111} + a_{1}b_{1}y_{1100} + a_{1}b_{1}y_{1101} + a_{1}b_{1}y_{1110} + a_{1}b_{1}y_{1111} = \Pr(Y_{a_0b_1}=y_1) $
\item $a_{0}b_{0}y_{0000} + a_{0}b_{0}y_{0001} + a_{0}b_{0}y_{0100} + a_{0}b_{0}y_{0101} + a_{0}b_{0}y_{1000} + a_{0}b_{0}y_{1001} + a_{0}b_{0}y_{1100} + a_{0}b_{0}y_{1101} + a_{0}b_{1}y_{0000} + a_{0}b_{1}y_{0001} + a_{0}b_{1}y_{0100} + a_{0}b_{1}y_{0101} + a_{0}b_{1}y_{1000} + a_{0}b_{1}y_{1001} + a_{0}b_{1}y_{1100} + a_{0}b_{1}y_{1101} + a_{1}b_{0}y_{0000} + a_{1}b_{0}y_{0001} + a_{1}b_{0}y_{0100} + a_{1}b_{0}y_{0101} + a_{1}b_{0}y_{1000} + a_{1}b_{0}y_{1001} + a_{1}b_{0}y_{1100} + a_{1}b_{0}y_{1101} + a_{1}b_{1}y_{0000} + a_{1}b_{1}y_{0001} + a_{1}b_{1}y_{0100} + a_{1}b_{1}y_{0101} + a_{1}b_{1}y_{1000} + a_{1}b_{1}y_{1001} + a_{1}b_{1}y_{1100} + a_{1}b_{1}y_{1101} = \Pr(Y_{a_1b_0}=y_0) $
\item $a_{0}b_{0}y_{0010} + a_{0}b_{0}y_{0011} + a_{0}b_{0}y_{0110} + a_{0}b_{0}y_{0111} + a_{0}b_{0}y_{1010} + a_{0}b_{0}y_{1011} + a_{0}b_{0}y_{1110} + a_{0}b_{0}y_{1111} + a_{0}b_{1}y_{0010} + a_{0}b_{1}y_{0011} + a_{0}b_{1}y_{0110} + a_{0}b_{1}y_{0111} + a_{0}b_{1}y_{1010} + a_{0}b_{1}y_{1011} + a_{0}b_{1}y_{1110} + a_{0}b_{1}y_{1111} + a_{1}b_{0}y_{0010} + a_{1}b_{0}y_{0011} + a_{1}b_{0}y_{0110} + a_{1}b_{0}y_{0111} + a_{1}b_{0}y_{1010} + a_{1}b_{0}y_{1011} + a_{1}b_{0}y_{1110} + a_{1}b_{0}y_{1111} + a_{1}b_{1}y_{0010} + a_{1}b_{1}y_{0011} + a_{1}b_{1}y_{0110} + a_{1}b_{1}y_{0111} + a_{1}b_{1}y_{1010} + a_{1}b_{1}y_{1011} + a_{1}b_{1}y_{1110} + a_{1}b_{1}y_{1111} =\Pr(Y_{a_1b_0}=y_1) $
\item $a_{0}b_{0}y_{0000} + a_{0}b_{0}y_{0010} + a_{0}b_{0}y_{0100} + a_{0}b_{0}y_{0110} + a_{0}b_{0}y_{1000} + a_{0}b_{0}y_{1010} + a_{0}b_{0}y_{1100} + a_{0}b_{0}y_{1110} + a_{0}b_{1}y_{0000} + a_{0}b_{1}y_{0010} + a_{0}b_{1}y_{0100} + a_{0}b_{1}y_{0110} + a_{0}b_{1}y_{1000} + a_{0}b_{1}y_{1010} + a_{0}b_{1}y_{1100} + a_{0}b_{1}y_{1110} + a_{1}b_{0}y_{0000} + a_{1}b_{0}y_{0010} + a_{1}b_{0}y_{0100} + a_{1}b_{0}y_{0110} + a_{1}b_{0}y_{1000} + a_{1}b_{0}y_{1010} + a_{1}b_{0}y_{1100} + a_{1}b_{0}y_{1110} + a_{1}b_{1}y_{0000} + a_{1}b_{1}y_{0010} + a_{1}b_{1}y_{0100} + a_{1}b_{1}y_{0110} + a_{1}b_{1}y_{1000} + a_{1}b_{1}y_{1010} + a_{1}b_{1}y_{1100} + a_{1}b_{1}y_{1110} = \Pr(Y_{a_1b_1}=y_0) $
\item $a_{0}b_{0}y_{0001} + a_{0}b_{0}y_{0011} + a_{0}b_{0}y_{0101} + a_{0}b_{0}y_{0111} + a_{0}b_{0}y_{1001} + a_{0}b_{0}y_{1011} + a_{0}b_{0}y_{1101} + a_{0}b_{0}y_{1111} + a_{0}b_{1}y_{0001} + a_{0}b_{1}y_{0011} + a_{0}b_{1}y_{0101} + a_{0}b_{1}y_{0111} + a_{0}b_{1}y_{1001} + a_{0}b_{1}y_{1011} + a_{0}b_{1}y_{1101} + a_{0}b_{1}y_{1111} + a_{1}b_{0}y_{0001} + a_{1}b_{0}y_{0011} + a_{1}b_{0}y_{0101} + a_{1}b_{0}y_{0111} + a_{1}b_{0}y_{1001} + a_{1}b_{0}y_{1011} + a_{1}b_{0}y_{1101} + a_{1}b_{0}y_{1111} + a_{1}b_{1}y_{0001} + a_{1}b_{1}y_{0011} + a_{1}b_{1}y_{0101} + a_{1}b_{1}y_{0111} + a_{1}b_{1}y_{1001} + a_{1}b_{1}y_{1011} + a_{1}b_{1}y_{1101} + a_{1}b_{1}y_{1111} =\Pr(Y_{a_1,b_1}=y_1) $
\end{itemize}
\item Monotonicity: \\
$a_0b_0y_{1000} + 2\cdot a_0b_0y_{1001} + a_0b_0y_{1100} + a_0b_0y_{1101} + a_0b_1y_{0100} + a_0b_1y_{0110} + a_0b_1y_{1100} + a_0b_1y_{1110} + a_1b_0y_{1000} + a_1b_0y_{1001} + a_1b_0y_{1100} + a_1b_0y_{1101} + a_1b_1y_{0100} + 2 \cdot a_1b_1y_{0110} + a_1b_1y_{1100} + a_1b_1y_{1110} + a_0b_0y_{1010} + a_0b_0y_{1011} + a_0b_1y_{1000} + a_0b_1y_{1001} + a_0b_1y_{1010} + a_0b_1y_{1011} + a_1b_0y_{0010} + a_1b_0y_{0110} + a_1b_0y_{1010} + a_1b_0y_{1110} + a_1b_1y_{0010} + a_1b_1y_{1010} = 0$
        \item Axioms of probability :
        \begin{itemize}
            \item         $\sum_{i = 0}^{1} \sum_{j = 0}^{1} \sum_{k = 0}^{1111} a_{i}b_{j}y_{k} = 1$ (Parameters sum to $1$) 
            \item $0 \leq a_{i}b_{j}y_{k}  \leq 1$, $\forall i \in \{0,1\}, \forall j \in \{0,1\}, \forall k \in \{0, 1, 10, 11, ..., 1111\}$
        \end{itemize}
    \end{itemize}

\item Objective functions:
\begin{itemize}
    \item Single treatment: \\
$a_{0}b_{0}y_{0010} +  a_{0}b_{0}y_{0011} +  a_{0}b_{0}y_{0110} +  a_{0}b_{0}y_{0111} +  a_{0}b_{0}y_{1010} +  a_{0}b_{0}y_{1011} +  a_{0}b_{0}y_{1110} +  a_{0}b_{0}y_{1111} +  a_{0}b_{1}y_{0001} +  a_{0}b_{1}y_{0011} +  a_{0}b_{1}y_{0101} +  a_{0}b_{1}y_{0111} +  a_{0}b_{1}y_{1001} +  a_{0}b_{1}y_{1011} +  a_{0}b_{1}y_{1101} +  a_{0}b_{1}y_{1111} +  a_{1}b_{0}y_{0010} +  a_{1}b_{0}y_{0011} +  a_{1}b_{0}y_{0110} +  a_{1}b_{0}y_{0111} +  a_{1}b_{0}y_{1010} +  a_{1}b_{0}y_{1011} +  a_{1}b_{0}y_{1110} +  a_{1}b_{0}y_{1111} +  a_{1}b_{1}y_{0001} +  a_{1}b_{1}y_{0011} +  a_{1}b_{1}y_{0101} +  a_{1}b_{1}y_{0111} +  a_{1}b_{1}y_{1001} +  a_{1}b_{1}y_{1011} +  a_{1}b_{1}y_{1101} +  a_{1}b_{1}y_{1111}$ = $\EX(Y_{a_1})$
\item ATE: \\
$a_0b_0y_{0010} +  a_0b_0y_{0011} +  a_0b_0y_{0110} +  a_0b_0y_{0111} +  a_0b_1y_{0001} +  a_0b_1y_{0011}+  a_0b_1y_{1001}+  a_0b_1y_{1011} +  a_1b_0y_{0010} +  a_1b_0y_{0011}+  a_1b_0y_{0110}+  a_1b_0y_{0111}+  a_1b_1y_{0001}+  a_1b_1y_{0011}+  a_1b_1y_{1001}+  a_1b_1y_{1011} - a_0b_0y_{1000} -  a_0b_0y_{1001} -  a_0b_0y_{1100} -  a_0b_0y_{1101} -  a_0b_1y_{0100} -  a_0b_1y_{0110} -  a_0b_1y_{1100} -  a_0b_1y_{1110} -  a_1b_0y_{1000} -  a_1b_0y_{1001} -  a_1b_0y_{1100} -  a_1b_0y_{1101} -  a_1b_1y_{0100} -  a_1b_1y_{0110} -  a_1b_1y_{1100} -  a_1b_1y_{1110} = \EX(Y_{a_1} - Y_{a_0})$
\end{itemize}
\end{itemize}

\end{document}